\documentclass[a4paper,fleqn]{cas-sc}

\usepackage[numbers]{natbib}
\usepackage{graphicx}%
\usepackage{multirow}%
\usepackage{amsmath,amssymb,amsfonts}%
\usepackage{amsthm}%
\usepackage{mathrsfs}%
\usepackage[title]{appendix}%
\usepackage{xcolor}%
\usepackage{textcomp}%
\usepackage{manyfoot}%
\usepackage{booktabs}%
\usepackage{algorithm}%
\usepackage{algorithmicx}%
\usepackage{algpseudocode}%
\usepackage{listings}%
\newtheorem{remark}{Remark}%

\usepackage[utf8]{inputenc}
\usepackage[T1]{fontenc}
\usepackage{comment}
\usepackage{color}
\usepackage{caption}
\usepackage{subcaption}

 \definecolor{green2}{rgb}{0, 0.5, 0}

\newcommand{\vecTheta}{\text{\boldmath$\theta$}}
\newcommand{\matTheta}{\mathbf{\Theta}}

\DeclareMathOperator*{\argmin}{arg\,min}

\def\tsc#1{\csdef{#1}{\textsc{\lowercase{#1}}\xspace}}
\tsc{WGM}
\tsc{QE}
\tsc{EP}
\tsc{PMS}
\tsc{BEC}
\tsc{DE}

\begin{document}
\let\WriteBookmarks\relax
\def\floatpagepagefraction{1}
\def\textpagefraction{.001}

\shorttitle{WLaSDI}

\shortauthors{A Tran et~al.}

\title [mode = title]{Weak-Form Latent Space Dynamics Identification}                      
\tnotemark[1]



%
\author[1]{April Tran}[auid=000,bioid=1,orcid=0009-0007-5319-7619]

\cormark[2]


\ead{chi.tran@colorado.edu}


\credit{Conceptualization of this study, Methodology, Software}

\affiliation[1]{organization={Department of Applied Mathematics},
    addressline={University of Colorado}, 
    city={Boulder},
    state={CO},
    postcode={80309-0526}, 
    country={USA}}

\author[2]{Xiaolong He}[orcid=0000-0002-5307-0681,type=author]
\ead{xiaolong.he@ansys.com}

\author[1]{Daniel A. Messenger}[orcid=0000-0002-8275-7888]
\ead{daniel.messenger@colorado.edu}

\credit{Data curation, Writing - Original draft preparation}

\affiliation[2]{organization={ANSYS Inc.},
    city={Livermore},
    state={CA},
    postcode={94551}, 
    country={USA}}

\author[3]{Youngsoo Choi}[orcid=0000-0001-8797-7970]
\ead{choi15@llnl.gov}

\affiliation[3]{organization={Center for Applied Scientific Computing},
    addressline={Lawrence Livermore National Laboratory}, 
    city={Livermore},
    state={CA},
    postcode={94550}, 
    country={USA}}

\author[1]{David M. Bortz}[orcid=0000-0003-1163-7317]
\ead{david.bortz@colorado.edu}
\cormark[1]
\cortext[cor1]{Corresponding author}
\cortext[cor2]{Principal corresponding author}



\begin{abstract}
Recent work in data-driven modeling has demonstrated that a weak formulation of model equations enhances the noise robustness of a wide range of computational methods. In this paper, we demonstrate the power of the weak form to enhance the LaSDI (Latent Space Dynamics Identification) algorithm, a recently developed data-driven reduced order modeling technique.  

We introduce a weak form-based version WLaSDI (Weak-form Latent Space Dynamics Identification).  WLaSDI first compresses data, then projects onto the test functions and learns the local latent space models. Notably, WLaSDI demonstrates significantly enhanced robustness to noise. With WLaSDI, the local latent space is obtained using weak-form equation learning techniques. Compared to the standard sparse identification of nonlinear dynamics (SINDy) used in LaSDI, the variance reduction of the weak form guarantees a robust and precise latent space recovery, hence allowing for a fast, robust, and accurate simulation.  We demonstrate the efficacy of WLaSDI vs. LaSDI on several common benchmark examples including viscid and inviscid Burgers', radial advection, and heat conduction. For instance, in the case of 1D inviscid Burgers' simulations with the addition of up to 100\% Gaussian white noise, the relative error remains consistently below 6\% for WLaSDI, while it can exceed 10,000\% for LaSDI. Similarly, for radial advection simulations, the relative errors stay below 15\% for WLaSDI, in stark contrast to the potential errors of up to 10,000\% with LaSDI.Moreover, speedups of several orders of magnitude can be obtained with WLaSDI. For example applying WLaSDI to 1D Burgers' yields a 140X speedup compared to the corresponding full order model.

Python code to reproduce the results in this work is available at (\url{https://github.com/MathBioCU/PyWSINDy\_ODE}) and (\url{https://github.com/MathBioCU/PyWLaSDI}).
\end{abstract}


\begin{highlights}
\item We introduce WLaSDI: Weak form Latent Space Dynamics Identification, a data-driven simulation framwork that leverages WENDy: Weak form Parameter Estimation.
\item  WLaSDI expands upon LaSDI: Latent Space Dynamics Identification in form and functionality.
\item We conduct a performance comparison between WLaSDI and LaSDI across several benchmark problems: viscid and inviscid Burgers', radial advection, and heat conduction. 
\item WLaSDI exhibits significantly greater robustness to noise compared to LaSDI. 
\end{highlights}

\maketitle

\section{Introduction}
Data-driven modeling harnesses the potential of  data to provide valuable insights and predictions with broad applications across various domains, such as uncertainty quantification, inverse problems, and design optimization. A significant achievement in the realm of data-driven modeling is the SINDy (Sparse Identification of Nonlinear Dynamics) algorithm, as described in the reference \cite{BruntonProctorKutz2016ProcNatlAcadSci}. It involves the formulation and discretization of system discovery problems, where nonlinear candidate basis functions are evaluated with given datasets. The candidate functions are then pruned using sparse regression methods to discover the model that best fits the data (in $L^2$ sense).

The first generation of the SINDy-based algorithms \cite{BruntonProctorKutz2016ProcNatlAcadSci,RudyBruntonProctorEtAl2017SciAdv} were highly sensitive to noise in the data.  Within the last few years, several groups have demonstrated that an integral transform of the proposed model offers significant robustness in the presence of noise.  Schaeffer and McCalla \cite{SchaefferMcCalla2017PhysRevE} illustrated that a pure integral formulation could be quite effective.  However, the consensus has emerged that a weak formulation with carefully chosen test functions offers superior performance and robustness.\footnote{The pysindy codebase \cite{deSilvaChampionQuadeEtAl2020JOSS,KaptanogludeSilvaFaselEtAl2022JOSS} includes a weak formulation and several recent publications that use SINDy acknowledge that using the weak form to manage noisy data \cite{BertsimasGurnee2023NonlinearDyn,BruntonKutz2023arXiv230317078,NicolaouHuoChenEtAl2023PhysRevResearch, KaptanogluHansenLoreEtAl2023PhysicsofPlasmas,KaptanogluZhangNicolaouEtAl2023NonlinearDyn}.}

In this paper, we contribute to this growing body of research by introducing an application of the weak form in the context of system discovery for data-driven reduced-order modeling (ROM). Through this application, we aim to showcase the efficacy of the weak form in comparison to the traditional strong form, emphasizing its ability to enhance robustness in the construction of reliable data-driven models in the presence of noise.

Reduced order modeling is inspired by the need to address the computational and mathematical challenges posed by simulating complex, high-dimensional systems in various scientific and engineering domains. To address this computational challenge, researchers have developed surrogate models to accelerate simulations while maintaining high accuracy. One notable method in this realm is the projection-based reduced order model (pROM), which approximates full-state fields through two families of compression techniques. Linear compression methods encompass well-known approaches such as proper orthogonal decomposition (POD) \cite{BerkoozHolmesLumley1993AnnuRevFluidMech, CheungChoiCopelandEtAl2023JournalofComputationalPhysics, ChoiBoncoraglioAndersonEtAl2020JournalofComputationalPhysics, ChoiBrownArrighiEtAl2021JournalofComputationalPhysics, ChoiCarlberg2019SIAMJSciComputa, ChoiCoombsAnderson2020SIAMJSciComput, ChoiOxberryWhiteEtAl2019arXiv190911320}, reduced basis \cite{RozzaHuynhPatera2008ArchComputatMethodsEng}, and balanced truncation \cite{SafonovChiang1989IEEETransAutomatContr}, while nonlinear compression methods make use of autoencoders (AE) \cite{KimChoiWidemannEtAl2020arXiv201107727, KimChoiWidemannEtAl2022JournalofComputationalPhysics, LeeCarlberg2020JournalofComputationalPhysics, MaulikLuschBalaprakash2021PhysFluids}. In this paper, we specifically explore non-intrusive pROM techniques that are purely data-driven. Interpolation techniques are commonly used to build nonlinear mappings that predict outputs based on inputs. These techniques encompass Gaussian processes \cite{TapiaKhairallahMatthewsEtAl2018IntJAdvManufTechnol, QianSeepersadJosephEtAl2006JMechDes}, radial basis functions \cite{DanielMarjavaaraStaffanLundstromGoelEtAl2007JFluidsEng, HuangWangYang2015}, Kriging \cite{HanGortz2012AIAAJournal, HanGortzZimmermann2013AerospaceScienceandTechnology}, and convolutional neural networks \cite{GuoLiIorio2016Proc22ndACMSIGKDDIntConfKnowlDiscovDataMin, ZhangSungMavris20182018AIAAASCEAHSASCStructStructDynMaterConf}. Neural networks are particularly popular due to their rich representation and feature extraction capabilities. However, they often lack interpretability and generalizability since they may require retraining for different output quantities

In this work, the ROM consists of so-called \emph{latent space} variables, which are low dimensional approximations of the full-order model. This class of methods involves first projecting (i.e., \emph{compressing}) the entire state field data into a reduced space followed by learning the ROM dynamics. There are two primary approaches to projection: linear and nonlinear. Linear compression, a widely used method, is achieved through the POD framework. In contrast, nonlinear compression is achieved using neural network autoencoders (AE). These compression techniques significantly reduce the spatial dimension of the data as well as simplify the data dynamics. 

The second step involves learning the latent space dynamics. One approach employs a latent space-time integrator to propagate solutions within the reduced space \cite{KimAzevedoThuereyEtAl2019ComputerGraphicsForum}. Alternatively, interpolation techniques such as artificial neural networks (ANNs) and radial basis functions (RBFs) are utilized to predict solutions for new parameter values \cite{KadeethumBallarinChoiEtAl2022AdvancesinWaterResources, FrescaDedeManzoni2021JSciComput}. One can also apply a linear multi-step neural network to predict and propagate latent space dynamics \cite{XieZhangWebster2019Mathematics}. However, it is worth noting that these methods often involve complex forms for latent space dynamics models, which can lead to a lack of interpretability. To tackle this challenge, the authors in \cite{ChampionLuschKutzEtAl2019arXiv190402107} employ the SINDy algorithm \cite{BruntonProctorKutz2016ProcNatlAcadSci} to uncover the governing equations that describe the latent dynamics. However, their approach falls short of achieving robust parametric modeling. This limitation arises because a single set of ordinary differential equations (ODEs) is insufficient to capture the full range of variations in the latent space caused by changes in system parameters. In response to this limitation, the LaSDI (Parametric Latent Space Dynamics Identification) algorithm, as introduced in \cite{FriesHeChoi2022ComputerMethodsinAppliedMechanicsandEngineering,HeChoiFriesEtAl2022arXiv220412005, he2022certified, BonnevilleChoiGhoshEtAl2024ComputMethodsApplMechEng}, offers an extension and parameterization of the concept. Additionally, it integrates SINDy to unveil the governing equations of the latent space. Nevertheless, when the input data is corrupted by noise, SINDy exhibits its limitations, struggling to accurately retrieve the system's dynamics. The weak form system discovery, on the other hand, performs well with noisy data due to the variance reduction nature of the integral. This reduction paves the way for a more robust and accurate latent space dynamics identification. 

In this paper, we present a novel approach known as Weak-form Latent Space Dynamics Identification (WLaSDI), which diverges from the use of SINDy in LaSDI. By replacing SINDy as the parameter estimator with the WENDy (weak form estimation of nonlinear dynamics) algorithm \cite{BortzMessengerDukic2023BullMathBiol}, WLaSDI offers significant advantages over the standard SINDy implementation found in LaSDI. Through a series of numerical experiments, we demonstrate the capabilities of our proposed framework, highlighting its accuracy, efficiency, and robustness. These experiments demonstrate the practicality and versatility of WLaSDI across various application scenarios.

This paper is structured as follows. In Section \ref{sec:background}, we delve into the essential background information for this study, which encompasses an exploration of the strong form LaSDI approach as outlined in Section \ref{subsec:lasdi}, as well as an examination of the weak-form equation learning and parameter estimation, discussed in Section \ref{subsec:weakform}. For a thorough understanding of the weak form latent space dynamics identification, we present a step-by-step breakdown in Section \ref{sec:wlasdi}. Then, the numerical results derived from our research are elaborated upon in Section \ref{sec:results}. Lastly, in Section \ref{sec:conclusion}, we provide the concluding remarks and discuss the potential future direction. 

\subsection{Mathematical notation}\label{sec:notation}
We employ uppercase bold font to represent matrix-valued variables, lowercase bold font for vector-valued variables, and lowercase non-bold font for scalar-valued variables.

\section{Background}\label{sec:background}

\subsection{Weak-form Equation Learning and Parameter Estimation}\label{subsec:weakform} Given a differential equation, multiplying both sides by a smooth compactly supported function and integrating creates a so-called \emph{weak} form of the equation.  This is a type of \emph{variational} form and has found broad use in the sciences and engineering.  Several mathematicians in the early part of the 20th century, including Galerkin and Courant, pioneered weak form-based computational methods leading to what is now called the \emph{finite element method}.
The basis for the modern mathematical analysis of the weak form started with the work of Laurent Schwartz, who won the 1950 Fields Medal for his Theory of Distributions, which recast the notion of a function acting on a point to one acting on an observation structure or \emph{test function} \cite{Schwartz1950}. This later led to Lax and Milgram \cite{LaxMilgram1955ContributionstotheTheoryofPartialDifferentialEquations} studying parabolic PDEs via convolving the equation with a test function and proving the uniqueness properties of the solution in a Hilbert space.
While the weak form has found considerable success as both a tool for theoretical analysis and computation, it has only recently found success in data-driven modeling-based parameter estimation and reduced order modeling. 

In this section, we discuss the algorithms for weak form-based parameter estimation, WENDy \cite{BortzMessengerDukic2023BullMathBiol}, which is a crucial part of WLaSDI. 
In this work, we focus on a first-order dynamical system in $D$ dimensions given by:
\begin{equation}
    \frac{d}{dt} \mathbf{u}(t)= \mathbf{f}(\mathbf{u}(t)), \quad \mathbf{u}(0) = \mathbf{u}_0 \in  \mathbb{R}^{1\times D}
  \label{eq:ode}
\end{equation}
where $\mathbf{f}:\mathbb{R}^{1\times D} \to \mathbb{R}^{1\times D}$. Note that the right side of \eqref{eq:ode} assumes that it is a linear combination of the features $f_j$ such that for $\mathbf{W} \in \mathbb{R}^{J \times D}$ and $\vecTheta = [f_{1}, f_{2}, ..., f_{J}] \in\mathbb{R}^{1\times J}, f_j: \mathbb{R}^{1 \times D} \to \mathbb{R}, j \in [J]$,  we have:
\begin{equation}
\mathbf{f}(\mathbf{u}(t)) = \vecTheta(\mathbf{u}(t)) \mathbf{W}.
\end{equation}

To convert to the weak form, we multiply by an absolutely continuous test function $\phi(t): \mathbb{R} \to \mathbb{R}$ and integrate over the time domain: 
\begin{equation*}
    \int_a^b \frac{d}{dt} \mathbf{u}(t)\phi(t)dt =  \int_a^b\mathbf{f}(\mathbf{u}(t))\phi(t)dt = \int_a^b\vecTheta(\mathbf{u}(t)) \mathbf{W}\phi(t)dt
\end{equation*}
If we choose $\phi$ compactly supported in $(a, b)$, with integration by parts, we have:
\begin{equation}
   -\int_a^b \frac{d}{dt}\phi(t)\mathbf{u}(t)dt = \int_a^b\vecTheta(\mathbf{u}(t)) \mathbf{W}\phi(t)dt
   \label{equation1}
\end{equation}

The trapezoidal rule is used to numerically discretize the integrals in  \eqref{equation1}. We presume that the system's state observations take place at a discrete set of time points, denoted as $\{t_{m}\}_{m=0}^{M}$ with a uniform step size of $\Delta t$ \footnote{Non-uniform step size is an option, although it requires slight adjustments to the implementation.}.
We denote 
\begin{equation*}
\begin{array}{ccc}
\centering
\mathbf{t}:=\left[\begin{array}{c}
t_0\\
\vdots\\
t_M
\end{array}\right], \quad & \mathbf{U}:=\left[\begin{array}{ccc}
         u_1(t_0) & \cdots & u_D(t_0) \\
         \vdots & \ddots & \vdots \\
         u_1(t_M) & \cdots & u_D(t_M)
     \end{array}\right],\quad  &  \matTheta(\mathbf{U}):=\left[\begin{array}{cccc}
        \boldsymbol{f}_1(\mathbf{U}) &  \boldsymbol{f}_2(\mathbf{U}) & \cdots & \boldsymbol{f}_J(\mathbf{U})
        \end{array}\right],
\end{array}
\label{eq:defnotation}
\end{equation*}
where $\matTheta:\mathbb{R}^{(M+1)\times D} \rightarrow \mathbb{R}^{(M+1)\times J}$ and $\boldsymbol{f}_j(\mathbf{U}):\mathbb{R}^{(M+1) \times D} \rightarrow \mathbb{R}^{M+1}$ is defined as $\boldsymbol{f}_j(\mathbf{U}):=\left[\begin{array}{ccc}
         f_j(\mathbf{u}(t_0)) & \cdots &
         f_j(\mathbf{u}(t_M))
     \end{array}\right]^T$.
Let $\frac{d}{dt}\phi_k = \phi_k'$, we introduce the quadrature matrices:
\begin{equation}\label{eq:quadrature_matrix}
\begin{array}{rl}
\mathbf{\Phi}_{km} =  \Delta t \phi_k(t_m), \quad \mathbf{\Phi} \in \mathbb{R}^{K\times (M+1)}\\
\mathbf{\dot{\Phi}}_{km} =  \Delta t \phi'_k(t_m), \quad \mathbf{\dot{\Phi}} \in \mathbb{R}^{K\times (M+1)} 
\end{array}
\end{equation}
And then define
\begin{equation}\label{eq:G_b}
\begin{array}{rl}
\mathbf{G} & :=\left[\mathbb{I}_D \otimes\mathbf{\Phi}\mathbf{\Theta(}\mathbf{U})\right]\in\mathbb{R}^{(KD) \times (JD) },\\
\mathbf{b} & :=-\mathsf{vec}(\dot{\mathbf{\Phi}}\mathbf{U})\in\mathbb{R}^{(KD)},
\end{array}
\end{equation}
where $\mathbb{I}_D$ is the identity and $\mathsf{vec}$ is the column-major vectorization operator. 
By solving the least squares problem, we identify a vector of parameters $\mathbf{w}\in \mathbb{R}^{(JD)}$ by minimizing the 2-norm of the residual
\begin{equation}
\underset{\mathbf{w}\in\mathbb{R}^{JD}}{\text{minimize}} \hspace{3pt} \left\Vert \mathbf{G}\mathbf{w}-\mathbf{b}\right\Vert _{2}^{2}
\label{eq:WENDy}
\end{equation}
Although ordinary least squares has been proven valuable to solve for $\mathbf{w}$, in cases with large-scale separation between the parameters or strong linearities, the estimated $\mathbf{w}$ could be biased. Equation \eqref{eq:WENDy} differs from the standard least square problem because $\mathbf{G}$ and $\mathbf{b}$ are both dependent on (noisy) measurement data $\mathbf{U}$.  Consequently, errors in the data affect both $\mathbf{G}$ and $\mathbf{b}$, which is known as the Errors-in-Variables problem in statistics (see \cite{VanHuffelLemmerling2002} for an overview).  Correctly solving for $\mathbf{w}$ thus necessitates a more sophisticated algorithm to effectively address the challenge. 

The WENDy algorithm \cite{BortzMessengerDukic2023BullMathBiol} 
 utilizes an Iteratively Reweighted Least Squares (IRLS) method, as detailed in \cite{Jorgensen2012EncyclopediaofEnvironmetrics}, to account for parameter-dependent covariance. The algorithm alternates between two key steps: the computation of $\mathbf{w}$ and the update of the covariance matrix for the linear system. At its core, WENDy focuses on linearizing the residual of the linear system, accounting for factors like measurement noise, integration errors, and numerical integration errors caused by the quadrature of the data in $\mathbf{U}$. The covariance is approximated at each iteration using a linearization. Thus WENDy minimizes the residual to obtain $\mathbf{w}$ through a generalized least squares approach. We direct the interested reader to \cite{BortzMessengerDukic2023BullMathBiol} for more information.

\subsection{LaSDI}\label{subsec:lasdi}
This section provides an overview of LaSDI: Latent Space Dynamics Identification \cite{FriesHeChoi2022ComputerMethodsinAppliedMechanicsandEngineering}, a novel reduced order modeling technique designed to significantly enhance accuracy within localized regions of the parameter space. LaSDI achieves this by identifying latent space dynamics in the form of local Ordinary Differential Equation (ODE) systems. These local ODE systems operate within their own trust regions, ensuring precision within their designated regions. Importantly, when the collective trust regions cover the entire parameter space, LaSDI delivers consistent accuracy, effectively creating a parametric model based on latent space learning. LaSDI's workflow encompasses four integral stages:

\begin{enumerate}
    \item \textbf{Data Generation}: Produce simulation data for the full order model (FOM), specifically generating parametric time-dependent solution data.
    \item \textbf{Compression}: Utilize compression techniques on the simulation data, employing either singular value decomposition or autoencoder methods to create latent space data.
    \item \textbf{Dynamics Identification}: Determine the governing equations that most closely align with the latent space data through Sparse Identification of Nonlinear Dynamics (SINDy) \cite{BruntonProctorKutz2016ProcNatlAcadSci}.
    \item \textbf{Prediction}: Once the governing equations of the latent space are known, it is used to predict the latent space solution at the testing point. This prediction can be decompressed to recover the full-state solution.
\end{enumerate}
In essence, LaSDI utilizes latent space learning to create a robust parametric model that excels in providing simulation accuracy tailored to specific areas of the parameter space. This approach not only ensures reliable accuracy across the parameter space but also offers computational efficiency compared to high-fidelity models.

\section{Weak-form LaSDI}\label{sec:wlasdi}
Similar to LaSDI \cite{FriesHeChoi2022ComputerMethodsinAppliedMechanicsandEngineering} we start by considering an $N_s$-dimensional full order model, characterized by the parameterized ODE:
\begin{equation}
    \frac{d\mathbf{u}}{dt} = \mathbf{h}(\mathbf{u}, t), \quad \mathbf{u}(0;\boldsymbol{\mu}) = \mathbf{u}_0(\boldsymbol{\mu})
    \label{eq:param_ode}
\end{equation}
where $t \in [0, T]$ represents time. The solution to Equation \eqref{eq:param_ode} is denoted as $\mathbf{u}(t;\boldsymbol{\mu})$, with $\mathbf{u}: [0, T] \times \mathcal{D} \rightarrow \mathbb{R}^{1 \times N_s}$. The right-hand side, $\mathbf{h}: \mathbb{R}^{1 \times N_s} \times [0, T] \rightarrow \mathbb{R}^{1 \times N_s}$, can be either linear or nonlinear in the first argument. The initial value is given by $\mathbf{u}_0: \mathcal{D} \rightarrow \mathbb{R}^{1 \times N_s}$, with $\boldsymbol{\mu} \in \mathcal{D}$ representing an element in the parameter space. We consider the case where this parameter only affects the initial conditions.

The spatial domain is denoted as $\Omega $ and can be discretized using various techniques such as finite difference or finite element methods. We assume a consistent time discretization with a uniform time step denoted by $\Delta t \in \mathbb{R}$. Throughout the paper, we use the following notation: $\mathbf{u}_n := \mathbf{u}(t_n; \boldsymbol{\mu})$. 


\begin{figure}[ht]
  \centering
  \label{fig:wlasdi_scheme}
  \includegraphics[width=\linewidth]{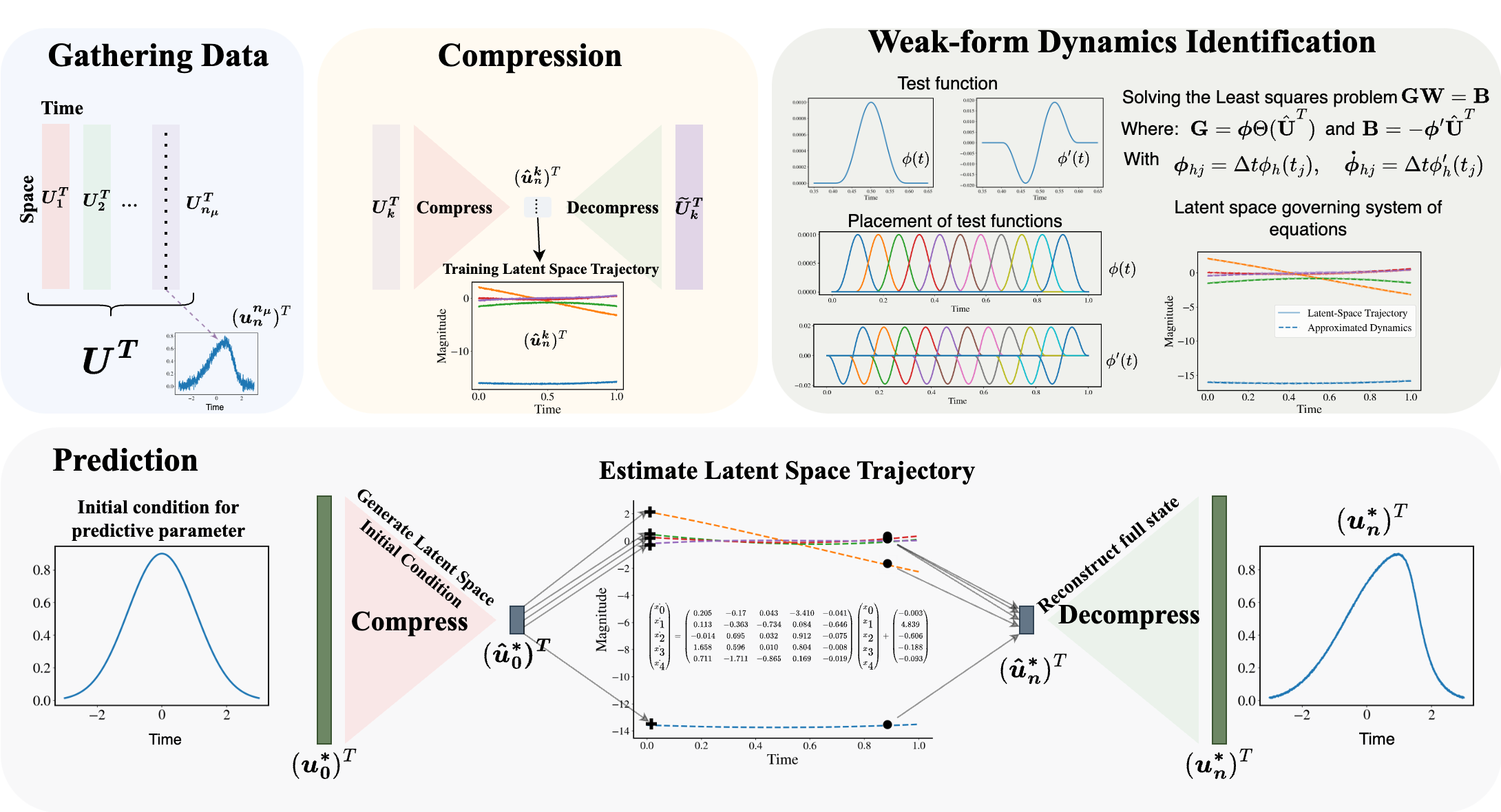}
  \caption{Diagram illustrating the WLaSDI algorithm's application to 1D Burgers' simulations, featuring four key steps: data gathering in section \ref{sec:wlasdi_data}, compression in section \ref{sec:wlasdi_compression}, weak form dynamics identification in section \ref{sec:wlasdi_DI}, and prediction in section \ref{sec:wlasdi_predict}.}
\end{figure}

\subsection{Data Gathering}\label{sec:wlasdi_data}
The initial stage of WLaSDI involves the generation or collection of simulation data for the full-order model, assumed to conform to Equation~\eqref{eq:param_ode}, by sampling the parameter space $\mathcal{D}$.  In contrast to LaSDI, WLaSDI accommodates noisy measurements. The sampling points within a training set $\mathcal{S}
\subset \mathcal{D}$ are denoted as $\boldsymbol{\mu}_k$, where $k \in \mathbb{N}(n_{\mu})$ and $n_{\mu}$ represents the number of sampling points in the training set.  \\
The solution at the $n$-th time step of Equation~\ref{eq:param_ode} with $\boldsymbol{\mu} = \boldsymbol{\mu}_k$ is represented by $\mathbf{u}_n^k \in \mathbb{R}^{1 \times N_s}$ and organized into a snapshot matrix 
$\mathbf{U}_k = \left[\begin{array}{cccc}
(\mathbf{u}_0^k)^T &  (\mathbf{u}_1^k)^T & \cdots & (\mathbf{u}_{N_t}^k)^T
\end{array}\right]^T \in \mathbb{R}^{(N_t+1) \times N_s}$. To compile the whole snapshot matrix $\mathbf{U} \in \mathbb{R}^{(N_t+1)n_{\mu} \times N_s }$, all individual snapshot matrices are concatenated as $\left[\begin{array}{cccc}
\mathbf{U}_0^T &  \mathbf{U}_1^T & \cdots & \mathbf{U}_{N_t}^T
\end{array}\right]^T$. \\
In Sections \ref{sec:wlasdi_compression} and \ref{sec:wlasdi_predict} presented below, we perform compression and decompression on the transpose of matrices $\mathbf{U}_k$ and $\mathbf{U}$, denoted as $\mathbf{U}_k^T$ and $\mathbf{U}^T$ respectively.

\subsection{Compression}\label{sec:wlasdi_compression} 
The second step of WLaSDI entails compressing the matrix $\mathbf{U}^T$ through the application of either linear or nonlinear compression techniques. The selection between these techniques depends on the behavior exhibited by the singular values of $\mathbf{U}^T$. If the singular values of $\mathbf{U}^T$ decay rapidly, we employ linear compression, specifically Proper Orthogonal Decomposition (POD). Conversely, when the singular values of $\mathbf{U}^T$ decay slowly, a nonlinear compression method is used. For a more comprehensive explanation of this decision-making process, please refer to Section 3.2 of \cite{FriesHeChoi2022ComputerMethodsinAppliedMechanicsandEngineering}.

Henceforth, the linear compression method in WLaSDI will be referred to as WLaSDI-LS (Linear Subspace), while the nonlinear compression method will be denoted as WLaSDI-NM (Nonlinear Manifold). We provide detailed explanations for both POD and our nonlinear compression method, shallow masked autoencoders, below.

\subsubsection{Proper Orthogonal Decomposition}\label{sec:pod}
The spatial basis generated by Proper Orthogonal Decomposition (POD) is a compact representation of $\mathbf{U}^T$ that reduces the projection error. This error represents the disparity between the initial snapshot matrix and the projected version on the subspace spanned by the basis $ \widehat{\mathbf{\Psi}}$: 
\begin{equation*}
    \widehat{\mathbf{\Psi}} := \argmin_{\mathbf{\Psi} \in \mathbb{R}^{N_s \times n_s}, \mathbf{\Psi}^T\mathbf{\Psi} = \mathbf{I}} \left\Vert\mathbf{U}^T - \mathbf{\Psi}\mathbf{\Psi}^T\mathbf{U}^T \right\Vert^2_F
\end{equation*}
To obtain the solution for (POD), we set $ \widehat{\mathbf{\Psi}} = \left[\begin{array}{cccc}\mathbf{v}_1 \quad \mathbf{v}_2 \quad \cdots \quad \mathbf{v}_{n_s} \end{array}\right]$ for $n_s < n_{\mu}(N_t + 1)$. We choose $n_s$ to be the dimension of the latent space. The vector $\mathbf{v}_k$ is the k-th column of the left singular matrix $\mathbf{V}$ from the Singular Value Decomposition (SVD), $\mathbf{U}^T = \mathbf{V}\mathbf{\Sigma}\mathbf{W}$. The generalized coordinate matrix $\widehat{\mathbf{U}}^T$ can be obtained by projecting the snapshot matrix $\mathbf{U}^T$ onto the subspace spanned by the column vectors of $\widehat{\mathbf{\Psi}}^T$. This results in a reduced snapshot matrix $\widehat{\mathbf{U}}^T \in \mathbb{R}^{n_s \times (N_t+1)n_{\mu}}$, i.e., 
\begin{equation*}
    \widehat{\mathbf{U}}^T :=  \widehat{\mathbf{\Psi}}^T\mathbf{U}^T
\end{equation*}
From the overall reduced snapshot matrix $\widehat{\mathbf{U}^T}$, the reduced snapshot matrix corresponding to a specific sampling point, $\boldsymbol{\mu}_k$, can be obtained by selecting the column vectors ranging from the $((k-1)(N_t + 1) + 1)$th to the $k(N_t + 1)$th in $\widehat{\mathbf{U}}^T$:
\begin{equation*}
    \widehat{\mathbf{U}}^T_k := \left[\begin{array}{cccc}(\hat{\mathbf{u}}_0^k)^T \quad (\hat{\mathbf{u}}_1^k)^T \quad \cdots \quad (\hat{\mathbf{u}}_{N_t}^k)^T \end{array}\right]
\end{equation*}
where $\widehat{\mathbf{u}}_j^k \in \mathbb{R}^{1 \times n_s}$ is the reduced coordinates at j-th time step for $\boldsymbol{\mu} = \boldsymbol{\mu}_k$. The matrix $\widehat{\mathbf{U}}^T_k \in \mathbf{R}^{n_s \times N_t}$ depicts the trajectory within the compressed subspace corresponding to the parameter value $\boldsymbol{\mu}_k$.

\subsubsection{Autoencoder}
Auto-encoders function as a nonlinear counterpart to POD. Two neural networks undergo training: one for the encoder $\mathcal{G}_{en}: \mathbb{R}^{N_s} \xrightarrow{} \mathbb{R}^{n_s} $ and another for the decoder: $\mathcal{G}_{de}: \mathbb{R}^{n_s} \xrightarrow{} \mathbb{R}^{N_s}$. The objective is to minimize the mean square error: 
\begin{equation}
    MSE(\mathbf{U^T} - \mathcal{G}_{de}(\mathcal{G}_{en}(\mathbf{U^T}))
    \label{eq:AE}
\end{equation}
Analogous to POD, we obtain reduced space data through $\widehat{\mathbf{U}}^T:= \mathcal{G}_{en}(\mathbf{U}^T) \in \mathbb{R}^{n_s \times (N_t+1)n_{\mu}}$ and extract the reduced snapshot matrix corresponding to a specific sampling point $\boldsymbol{\mu}_k$ by selecting the column vectors ranging from the $((k-1)(N_t + 1) + 1)$th to the $k(N_t + 1)$th in $\widehat{\mathbf{U}}^T$:
\begin{equation*}
    \widehat{\mathbf{U}}^T_k := \left[\begin{array}{cccc}(\hat{\mathbf{u}}_0^k)^T \quad (\hat{\mathbf{u}}_1^k)^T \quad \cdots \quad (\hat{\mathbf{u}}_{N_t}^k)^T \end{array}\right]
\end{equation*}
where $\widehat{\mathbf{u}}_j^k \in \mathbb{R}^{1 \times n_s}$ is the reduced coordinates at j-th time step for $\boldsymbol{\mu} = \boldsymbol{\mu}_k$.
In this paper, we have employed a shallow masked network for our autoencoder architecture, whose detail can be found in \cite{KimChoiWidemannEtAl2022JournalofComputationalPhysics, KimChoiWidemannEtAl2022JournalofComputationalPhysics}. Nevertheless, alternative advanced neural network architectures, such as deep convolutional neural networks \cite{lecun2015deep}, can also be applied.\\
\begin{remark}
The dimension of the latent space, denoted as $n_s$, plays a crucial role in both auto-encoder training and POD. If $n_s$
is chosen too small, the compression-decompression process may result in significant data loss, leading to a decrease in reconstruction accuracy. Conversely, opting for a large $n_s$ could introduce intricate dynamics in the latent space, posing challenges for subsequent dynamics learning, as discussed in the next subsection. Therefore, the choice of $n_s$ must strike a balance, aiming to minimize dimensionality while maintaining the accuracy of the reconstructed snapshot matrix.
\end{remark}
\subsection{Weak form Dynamics Identification }\label{sec:wlasdi_DI}

In this section, we demonstrate the weak form dynamics identification (DI). Our objective is to discover the latent space representation of 
$\dot{\hat{\mathbf{u}}}(t) = \mathbf{f}({\hat{\mathbf{u}}}(t))$ that best fits the discrete trajectory data in a least-squares sense. Note that $\hat{\mathbf{u}}: [0, T] \times \mathcal{D} \rightarrow \mathbb{R}^{1\times n_s}$ and $\mathbf{f}: \mathbb{R}^{1\times n_s} \rightarrow \mathbb{R}^{1\times n_s}$. Suppose the right-hand side can be written as 
$\mathbf{f}(\hat{\mathbf{u}}(t)) = \vecTheta(\mathbf{\hat{u}}(t)) \mathbf{W}$, where $\mathbf{W} \in \mathbb{R}^{M \times n_s}$ and $\vecTheta = [f_{1}, f_{2}, ..., f_{M}]\in\mathbb{R}^{1\times M}$. We denote a family of trial functions using the notations $ f_m: \mathbb{R}^{1 \times n_s} \to \mathbb{R}$, where $m \in [M]$. The functions $f_m$ may be linear or nonlinear.

Let $\widehat{\mathbf{U}}_k$ $\in \mathbb{R}^{(N_t+1) \times n_s}$ represent the latent space trajectory data associated with the parameter $\boldsymbol{\mu}_k$ and $\mathbf{t} = \{t_j\}_{j = 0}^{N_t}$ denote the time points. The trajectory data $\widehat{\mathbf{U}}_k$ takes the form: 
\begin{equation*}
    \widehat{\mathbf{U}}_k = \begin{bmatrix}
\hat{\mathbf{u}}_0  \\
\hat{\mathbf{u}}_1 \\
 \vdots \\
\hat{\mathbf{u}}_{N_t}
\end{bmatrix}
=  \begin{bmatrix}
\hat{u}_0(t_0) &\hat{u}_1(t_0)& \cdots & \hat{u}_{n_s}(t_0)  \\
\hat{u}_0(t_1) & \hat{u}_1(t_1) &\cdots & \hat{u}_{n_s}(t_1)  \\
 \vdots & \vdots & \ddots & \vdots \\
\hat{u}_{0}(t_{N_t}) & \hat{u}_{1}(t_{N_t}) & \cdots & \hat{u}_{n_s}(t_{N_t}). 
\end{bmatrix}
\end{equation*}
with a set of trial functions, $\{f_m\}_{m = 1}^M$. The trial function $f_m$ might be a constant, polynomial, trigonometric, logarithmic, exponential function, or composition of these nonlinear functions. 
We articulate the general form of the weak form dynamics identification by considering a family of test functions $\{\phi_h\}_{h = 1}^H$, where H is the number of test functions. The five steps for the weak form latent space dynamics identification are given as follows: 
\begin{enumerate}
    \item We broaden the scope of the trial function $\boldsymbol{f}_m$ by allowing it to accept data matrices as input, e.g., $\boldsymbol{f}_m(\widehat{\mathbf{U}}_k):=\left[\begin{array}{cccc}
         f_m(\mathbf{u}_0^k) & f_m(\mathbf{u}_1^k)&
         \cdots  &
         f_m(\mathbf{u}_{N_t}^k)
     \end{array}\right]^T\in\mathbb{R}^{N_t+1}$.
     We then construct the matrix of trial function: $\matTheta( \widehat{\mathbf{U}}_k ) =   \left[\begin{array}{cccc}\boldsymbol{f}_1(\widehat{\mathbf{U}}_k) & \boldsymbol{f}_2(\widehat{\mathbf{U}}_k) & \cdots & \boldsymbol{f}_M(\widehat{\mathbf{U}}_k) \end{array}\right] in\mathbb{R}^{(N_t+1)\times M}$. 
    
    \item Construct the integration matrices $ \mathbf{\Phi}, \mathbf{\dot{\Phi}}$ as in Eq.~\eqref{eq:quadrature_matrix}:
    \begin{equation*}
        \mathbf{\Phi}_{hj} = \Delta t \phi_h(t_j), \quad \mathbf{\dot{\Phi}}_{hj} = \Delta t \phi'_h(t_j)
    \end{equation*}
    \item Compute $\mathbf{G_k}:=\left[\mathbb{I}_{n_s} \otimes\mathbf{\Phi}\matTheta(\widehat{\mathbf{U}}_k)\right]\in\mathbb{R}^{(H n_s) \times (M n_s) }$ and $\mathbf{b}_k :=-\mathsf{vec}(\dot{\mathbf{\Phi}}\widehat{\mathbf{U}}_k)\in\mathbb{R}^{(Hn_s)}$ as in Eq.~\eqref{eq:G_b}. 
    
    \item Solve the least squares problem $\mathbf{G}_k \mathbf{w}_k = \mathbf{b}_k$  as in Eq.~\eqref{eq:WENDy}. 
    \item Reshape the vector $\mathbf{w}_k \in \mathbb{R}^{Mn_s}$ into $\mathbf{W}_k \in \mathbb{R}^{M \times n_s}$ such that $\mathbf{w}_k = \mathsf{vec}(\mathbf{W}_k)$ and we have the governing equation for the latent space dynamics: $\dot{\hat{\mathbf{u}}}_k = \mathbf{f}\left ({\hat{\mathbf{u}}_k}\right ) = \vecTheta\left( \hat{\mathbf{u}}_k \right)\mathbf{W}_k$. 
\end{enumerate}
In this paper, we employ two distinct families of compactly supported test functions, selected based on the specific problem context.  For the Burgers' and the heat conduction examples (refer to Sections \ref{sec:1DB}, \ref{sec:2DB}, and \ref{sec:Diff}), we use piecewise polynomials whose supports are chosen to match the spectrum of the data. Additional details on these test functions are available in Section 2.4 of \cite{MessengerBortz2021MultiscaleModelSimul}. For the radial advection examples described in Section \ref{sec:RA}, we utilize orthogonal test functions generated from a set of multiscale $C^\infty$ bump functions. Further elaboration on this approach is available in Section 2.3 of \cite{BortzMessengerDukic2023BullMathBiol}.

\begin{figure}[ht]
    \centering
    \includegraphics[width=0.5\linewidth]{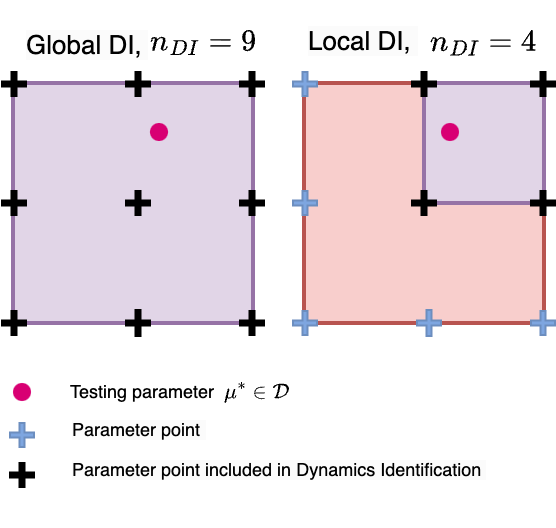}
    \caption{Illustrating Local vs Global DI: In a 2-dimensional parameter space with nine uniformly spaced training data points, we demonstrate the DI regions for an arbitrary parameter point $\mu^{*} \in \mathcal{D}$ denoted by the pink dot. Two distinct  DI regions are highlighted in purple, one for $n_{DI} = 9$ (global DI) and another for $n_{DI} = 4$ (local DI). The black crosses denote the parameter points covered by the DI regions.} 
    \label{fig:global_vs_local}
\end{figure}

After solving the least-squares problem $\mathbf{G}_k \mathbf{w}_k = \mathbf{b}_k$ for a single sample point $\boldsymbol{\mu}_k \in \mathcal{S}$, we establish a local DI. Expanding on this, we can improve the local DI by combining multiple snapshot data matrices $\mathbf{U}^T_k$ known as \textbf{region-based} DI. Conversely, when we employ the entire snapshot matrix $\mathbf{U}^T$ encompassing all sample points $\boldsymbol{\mu}_k \in \mathcal{S}$, we refer to it as \textbf{global} DI. Furthermore, another option is to interpolate the coefficients $\mathbf{w}_k$ for each $\boldsymbol{\mu}_k$, resulting in \textbf{interpolated} DI. We provide a brief overview of each type of DI below.  For more detailed information, please refer to \cite{FriesHeChoi2022ComputerMethodsinAppliedMechanicsandEngineering}. The differences between global DI and local DI are demonstrated in Figure \ref{fig:global_vs_local}. 
\subsubsection{Global Dynamics Identifiation} 
The implementation of Global DI is straightforward: we apply the weak form dynamics identification algorithm described earlier using the complete snapshot matrix 
$\mathbf{U}^T$ instead of $\mathbf{U}_k^T$. This DI model, which covers the entire parameter space, can be advantageous when latent space dynamics remain relatively stable across different parameter values. This approach is particularly effective for smaller parameter spaces. However, for high dimensional parameter spaces, significant variations in latent space dynamics may arise.

\subsubsection{Region-based Local Dynamics Identification} 
When we assume that closely spaced points in the parameter space $\mathcal{S}$ exhibit similar behavior, we can construct a local model that performs effectively within this specific region. The implementation of this region-based local DI closely mirrors that of the global DI. Instead of using either the entire snapshot matrix $\mathbf{U}^T$ or the snapshot matrix $\mathbf{U}^T_k$ from a single point $\boldsymbol{\mu}_k$, we concatenate $\mathbf{U}^T_k$ matrices of multiple points in a local region to create a local snapshot matrix $\widetilde{\mathbf{U}}^T$. Subsequently, we apply the algorithm described above to $\widetilde{\mathbf{U}}^T$ to derive the governing equation for latent space dynamics. The selection of the local region, represented by the set of points $\boldsymbol{\mu}_k$, can be made such that these points are in proximity to the testing point $\boldsymbol{\mu}_k^*$ that we aim to simulate.

\subsubsection{Interpolated Local Dynamics Identification}
We can determine the governing dynamics of a larger region by interpolating the governing equation derived for a smaller region contained within the larger one. In essence, we interpolate the coefficients $\mathbf{w}_k$ obtained for each $\boldsymbol{\mu}_k$. This paper employs various interpolation schemes such as radial basis functions or linear bivariate splines. It remains imperative to underscore that when choosing the local region, denoted by the set of points $\boldsymbol{\mu}_k$, these points should be in close proximity to the testing point $\boldsymbol{\mu}_k^*$ that is the target of our simulation.

\subsection{Prediction}\label{sec:wlasdi_predict}
After the Dynamics Identification step has been established through local, global, or interpolated DIs, the identified governing equation can be solved to predict new latent space dynamics trajectories for a given parameter value $\boldsymbol{\mu}^*$ within the parameter space $\mathcal{D}$. Accurate forecasting for this parameter $\boldsymbol{\mu}^*$ necessitates an appropriate initial condition $\hat{\mathbf{u}}_0^*$ in the latent space. This initial condition can be derived by compressing the initial condition $(\mathbf{u}^*_0)^T$ from the full-order model corresponding to the parameter value $\boldsymbol{\mu}^*$. Two compression techniques can be utilized, as mentioned earlier, i.e., linear or nonlinear compressions:
\begin{equation*}
(\hat{\mathbf{u}}_0^*)^T = \widehat{\mathbf{\Psi}}^T(\mathbf{u}^*_0)^T \quad \text{or} \quad \hat{\mathbf{u}}_0^* = \mathcal{G}_{en}((\mathbf{u}^*_0)^T)
\end{equation*}
Subsequently, this initial condition $(\hat{\mathbf{u}}_0^*)^T$ is used to generate the latent space trajectory $(\hat{\mathbf{u}}_n^*)^T$ by numerically solving the ODEs that are identified in DI phase. This trajectory can then be decompressed to recover the solution prediction in full-order space. When employing linear compression, the approximated full-state trajectories $(\mathbf{u}^*_n)^T$ are recovered using $(\mathbf{u}^*_n)^T = \widehat{\mathbf{\Psi}}(\hat{\mathbf{u}}_n^*)^T$. Likewise, for nonlinear compression, the decompression process is $(\mathbf{u}^*_n)^T = \mathcal{G}_{de}((\hat{\mathbf{u}}_n^*)^T)$. 


\section{Numerical Results}\label{sec:results}
In this section, we present a performance comparison between WLaSDI and LaSDI with and without the presence of noise. From this point forward, we will refer to WLaSDI with linear compression as WLaSDI-LS, where 'LS' denotes linear subspace. Similarly, WLaSDI with nonlinear compression will be denoted as WLaSDI-NM, where 'NM' signifies nonlinear manifold. We employ the same set of examples as detailed in \cite{FriesHeChoi2022ComputerMethodsinAppliedMechanicsandEngineering}. Specifically, we investigate four distinct examples, including the 1D inviscid Burgers' equations, the 2D viscous Burgers' equation, heat conduction, and the radial advection problem. Below, we present the governing equations and relevant parameters for each example in Table \ref{table:eqs}.
\begin{table}[h!]
\centering
 \begin{tabular}{||c |c |c |c | c||} 
 \hline \textbf{Equation} & \textbf{Initial Condition} & \textbf{Domain/Boundary Condition} & \textbf{Discretization}\footnotemark \\ [2ex] 
 \hline\hline
\begin{tabular}{@{}c@{}} \textbf{1D Burgers'} \\ $u_t = -uu_x$ \end{tabular} & \begin{tabular}{@{}c@{}}$u(0, x; a,w) = a \cdot \exp(\frac{-x^2}{w})$ \\ $a \in [0.7, 0.9]$ \\ $w \in [0.9, 1.1]$ \end{tabular} & \begin{tabular}{@{}c@{}}$\Omega = [-3, 3]$\\ $ t \in [0, 1]$ \\ $u(3, t) = u(-3, t) = 0$ \end{tabular} & \begin{tabular}{@{}c@{}}$\Delta x = 6/1000$\\ $\Delta t = 1/1000$ \end{tabular}\\ [4ex] 
 \hline
 \begin{tabular}{@{}c@{}} \textbf{2D Burgers'} \\ $\mathbf{u}_t = - \mathbf{u}\cdot \nabla\mathbf{u} + \frac{1}{10000} \Delta \mathbf{u}$ \end{tabular} &   \begin{tabular}{@{}c@{}}$\mathbf{u}(0, \mathbf{x}; a,w) = \begin{bmatrix}
a \cdot \exp(-||\mathbf{x}||^2_2/w) \\ a \cdot \exp(-||\mathbf{x}||^2_2/w) \end{bmatrix}$ \\ $a \in [0.7, 0.9]$ \\ $w \in [0.9, 1.1]$ \end{tabular} & \begin{tabular}{@{}c@{}}$\Omega = [-3, 3] \times [-3, 3]$\\ $ t \in [0, 2]$ \\ $\mathbf{u}(\mathbf{x}, t) = \mathbf{0}$ on $\partial \Omega$ \end{tabular}  & \begin{tabular}{@{}c@{}}$\Delta x = \Delta y = 1/10$\\ $\Delta t = 2/1500$ \end{tabular}\\ [5ex] 
\hline
\begin{tabular}{@{}c@{}} \textbf{Heat Conduction} \\  $u_t = \nabla \cdot (1 + u)\nabla u $ \end{tabular} & \begin{tabular}{@{}c@{}}$u(0, \mathbf{x}; a, \omega) = a \cdot sin(\omega(x_1 + x_2)) + a$ \\ $a \in [1.8, 2.2]$ \\ $\omega \in [0.2, 5.0]$ \end{tabular} & \begin{tabular}{@{}c@{}}$\Omega = [0, 1] \times [0, 1]$\\ $ t \in [0, 1]$ \\ $\frac{\partial u}{\partial n} = 0$ on $\partial \Omega$ \end{tabular}  & $\Delta t = 0.01$\\ [4ex] 
\hline
\begin{tabular}{@{}c@{}} \textbf{Radial Advection}\\  $u_t = - \mathbf{v} \cdot \nabla u$ \\ $\mathbf{v}  = \frac{\pi}{2}d \begin{bmatrix} 
x_2 \\-x_1 \end{bmatrix}$ \\  $d = ((1 - x_0^2)(1 - x_1^2))^2$\end{tabular}  &   \begin{tabular}{@{}c@{}}$u(0, \mathbf{x}; \omega) = sin(\pi\omega x_1)sin(\pi \omega x_2)$ \\ $\omega \in [0.6, 1.0]$ \\ or \\ $\omega \in [0.6, 1.4]$ \end{tabular} &\begin{tabular}{@{}c@{}}$\Omega = [-1, 1] \times [-1, 1]$\\ $ t \in [0, 3]$ \\ $u(\mathbf{x}, t) = 0$ on $\partial \Omega$ \end{tabular} & $\Delta t = 0.0025$\\ [0.5ex] 
 \hline
 \end{tabular}
 \caption{Equations used in numerical experiments}
 \label{table:eqs}
\end{table}

We adopt the same sampling approach for our full-order models as described in \cite{FriesHeChoi2022ComputerMethodsinAppliedMechanicsandEngineering}. In the case of both the 1D and 2D Burgers' equations, we apply finite differences to construct our full-order models. Further information regarding the discretization process can be found in Table \ref{table:eqs}. To be more specific, we employ the backward difference scheme for the first-order spatial derivative and the central difference scheme for the second-order spatial derivative. Additionally, we utilize the implicit backward Euler's method for time integration.

For both the heat conduction and radial advection equations, we employ the finite element method. Specifically, in the case of nonlinear heat conduction, the spatial domain is discretized using the second-order finite element approach, initially consisting of uniform $64 \times 64$ squares. These squares are subsequently subdivided into 8192 triangular elements. A uniform time step of $0.01$ is utilized, along with the SDIRK3 implicit L-stable time integrator. 
The temperature field-dependent conductivity coefficient is linearized by utilizing the temperature field from the preceding time step\footnote{The complete source code for the full-order model pertaining to the nonlinear heat conduction problem is available at https://github.com/mfem/mfem/blob/master/examples/ex16.cpp}.

For the radial advection problem, the full-order model employs the third-order discontinuous finite element method to partition the spatial domain. This domain is represented as a square mesh with periodic boundary conditions, divided into $24 \times 24$ square finite elements. Subsequently, the finite element data is interpolated to create a uniform $64 \times 64$ grid across the spatial domain. A uniform time step of $0.0025$ is applied using the fourth-order explicit Runge-Kutta method\footnote{The complete source code for the full-order model addressing the radial advection problem is accessible at https://github.com/mfem/mfem/blob/master/examples/ex9.cpp}.

\begin{table}[h!]
\centering
 \begin{tabular}{||c |c |c ||} 
 \hline \textbf{Equation} & \textbf{Training Set} & \textbf{Testing set}\\ [2ex] 
 \hline\hline
 \textbf{1D Burgers'} &  \begin{tabular}{@{}c@{}} $a \in [0.75, 0.85] $, $w \in [0.95, 1.05]$ \\ or \\ $a \in [0,7, 0,8, 0.9] $,$ w \in [0.9, 1.0, 1.1]$\end{tabular} & \begin{tabular}{@{}c@{}} $a = 0.85$ , $w = 0.95$ \\ or \\ $a \in [0.7, 0.71, \cdots, 0.9] $, $w \in [0.9, 0.91,  \cdots, 1.1]$\end{tabular} \\[4ex]
 \hline
 \textbf{2D Burgers'} &  \begin{tabular}{@{}c@{}} $a \in [0,7, 0,8, 0.9] $\\ $w \in [0.9, 1.0, 1.1]$ \end{tabular} & \begin{tabular}{@{}c@{}} $a = 0.85 $ \\ $w = 0.95$ \end{tabular} \\[3ex]
 \hline
 \textbf{Heat Conduction}&  \begin{tabular}{@{}c@{}} $a \in [1.8, 2.0. 2.2] $\\ $w \in [1.8, 2.0. 2.2] $ \end{tabular} & \begin{tabular}{@{}c@{}} $a = 1.85$ \\ $w = 2.05$ \end{tabular} \\[3ex]
 \hline
\textbf{Radial Advection} &  \begin{tabular}{@{}c@{}} $a \in [1.0,1.05, 1.1, 1.15, 1.2, 1.25, 1.3, 1.35, 1.4] $\\ $w \in [1.0]$ \end{tabular} & \begin{tabular}{@{}c@{}} $a = 1.23$ \\ $w = 1.0$ \end{tabular} \\[3ex]
 \hline
 \end{tabular}
 \caption{Training and testing parameter values}
 \label{table:params}
\end{table}
\subsection{1D Inviscid Burgers'}\label{sec:1DB}
In this section, we conduct a comparative analysis of LaSDI and WLaSDI for simulating the 1D Burgers' equations under both noisy and noise-free conditions. The 1D Burgers' equation, along with its associated domain, is detailed in Table \ref{table:eqs}. Various initial conditions for the equation are utilized as training data, i.e., $u(0, x; a,w) = a \cdot \exp(\frac{-x^2}{w})$ with different values for $a$ and $w$. Subsequently, LaSDI and WLaSDI are employed to simulate a distinct set of initial conditions referred to as testing points. Comprehensive information regarding the training and testing datasets can be found in Table \ref{table:params}. 

\begin{figure} 
\centering
\includegraphics[width=0.4\linewidth]{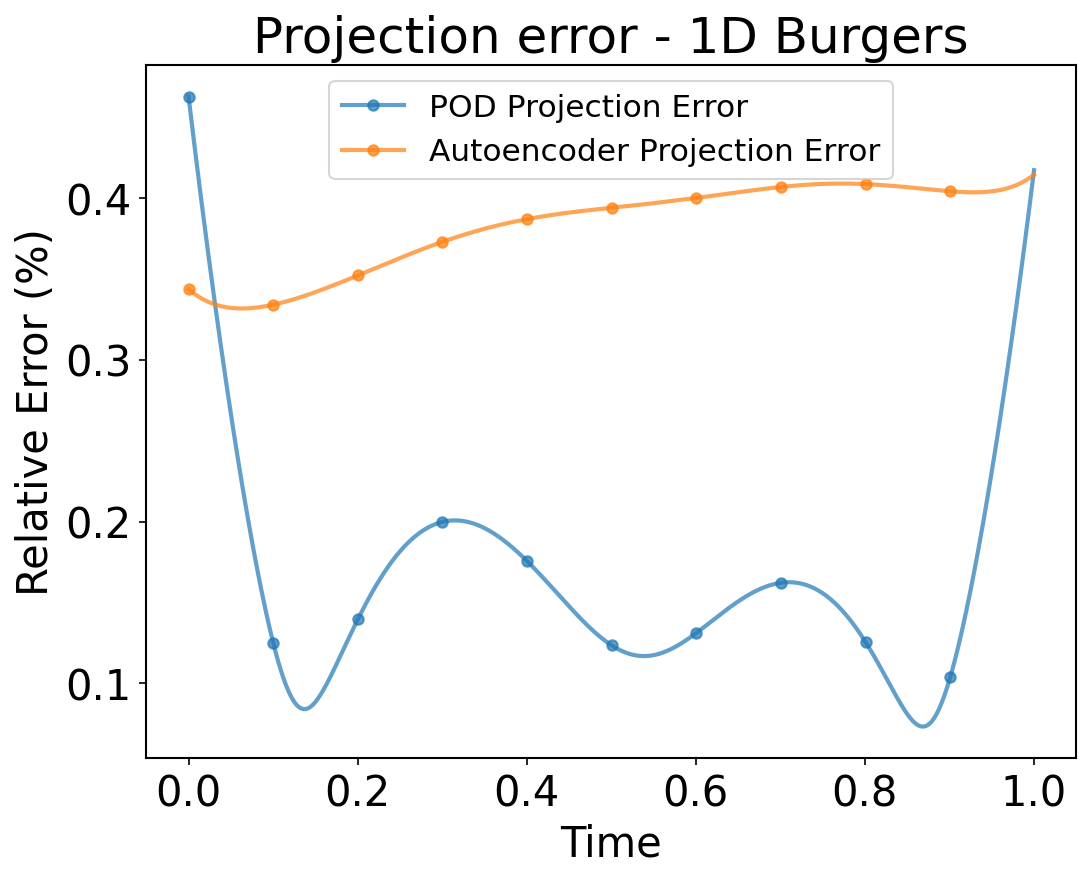}
    \caption{Projection errors of the noise-free 1D Burgers' equation resulting from both linear and nonlinear compression methods. Projection error quantifies the comparison between the data and the compressed-decompressed representation.}
    \label{fig:1DB_prj}
\end{figure}

\begin{figure}
    \centering
    \begin{subfigure}[t]{0.4\textwidth}
        \centering
        \includegraphics[width=\linewidth]{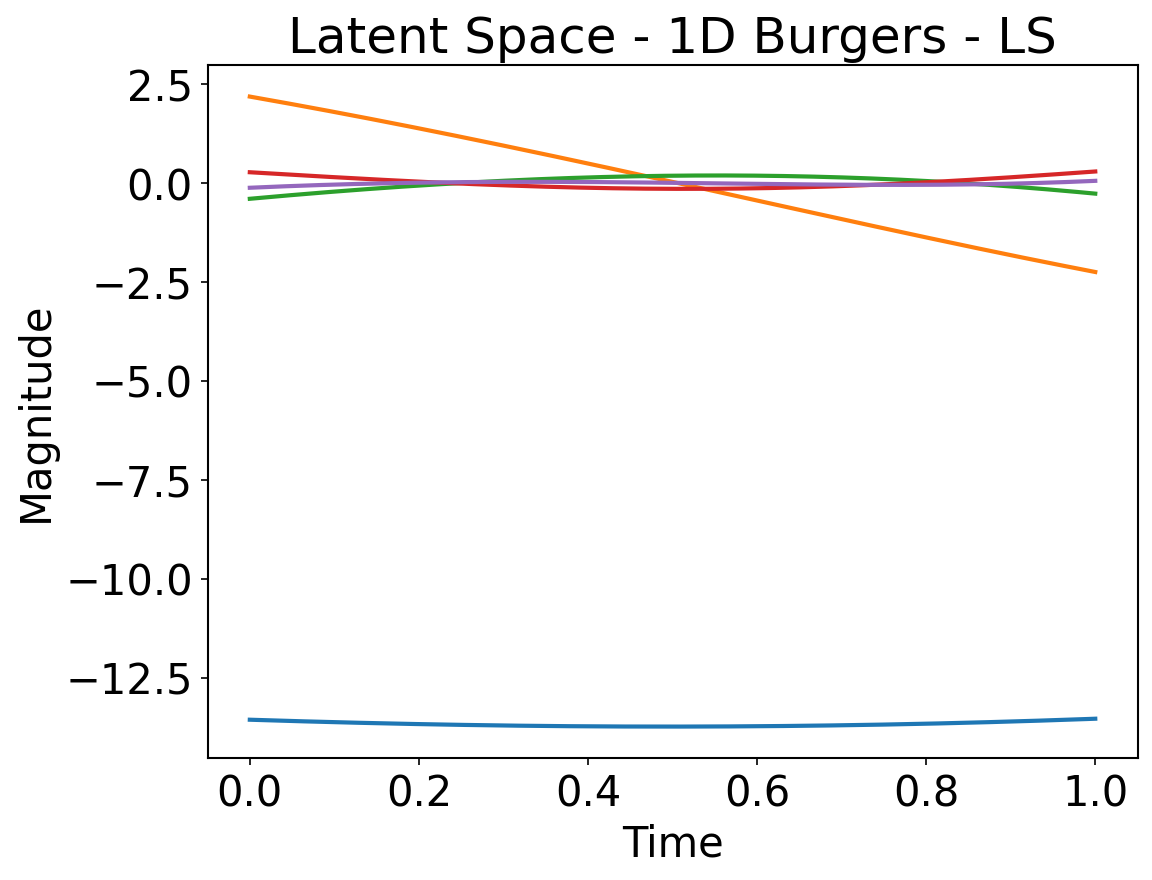}
        \caption{With linear compression}
    \end{subfigure}
    \begin{subfigure}[t]{0.4\textwidth}
        \centering
        \includegraphics[width=\linewidth]{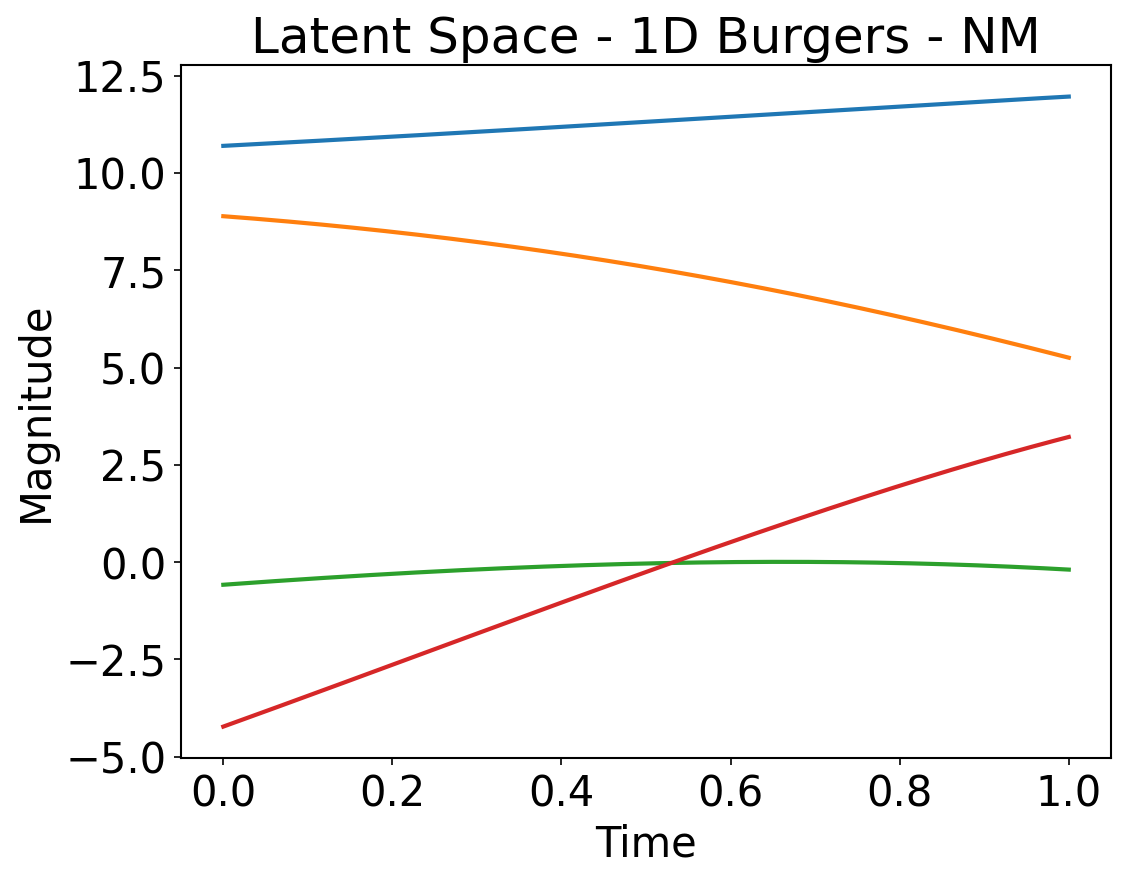}
        \caption{With nonlinear compression}
    \end{subfigure}
    \caption{Latent space trajectories of the noise-free 1D Burgers' equation obtained through both linear and nonlinear compression.}
    \label{fig:1DB_latent}
\end{figure}

In Figure \ref{fig:1DB_prj}, we present both the projection errors 
resulting from linear and nonlinear compression while in Figure \ref{fig:1DB_latent} we show their respective latent space trajectories. We opted for a latent space dimension of 4 for nonlinear compression and 5 for linear compression in alignment with LaSDI. When employing the POD, we analyze the cumulative sum of singular values and truncate to retain 99\% of the energy. In the case of the autoencoder, a latent space dimension of 4 was chosen, preserving 98\% of the energy. The projection error quantifies the disparity between the original data and their compressed-decompressed representations and is given by $E_p(\mathbf{U}) = \frac{\left\Vert \mathbf{U} - \mathcal{E}_{de}(\mathcal{E}_{en}(\mathbf{U})\right\Vert_2}{\left\Vert\mathbf{U}\right\Vert_2} $ where $\mathcal{E}_{en}$ and $\mathcal{E}_{de}$ are the encoder and decoder operator, respectively.

\begin{figure}
    \centering
    \begin{subfigure}[t]{0.4\textwidth} 
        \centering
        \includegraphics[width=\linewidth]{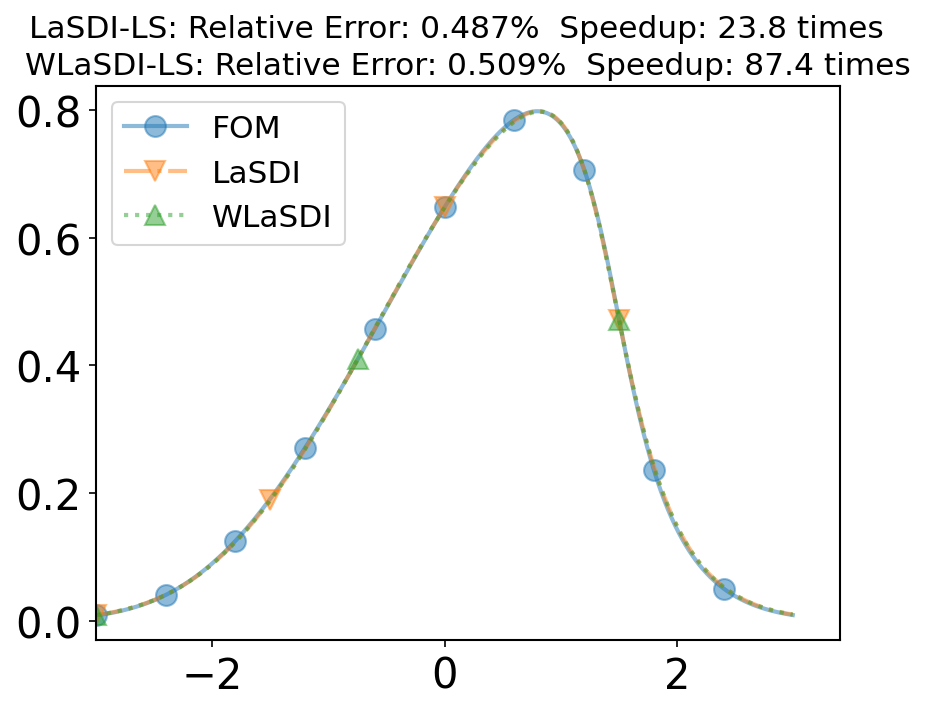}
        \caption{Using linear compression}
    \end{subfigure}
    \begin{subfigure}[t]{0.4\textwidth}
        \centering
        \includegraphics[width=\linewidth]{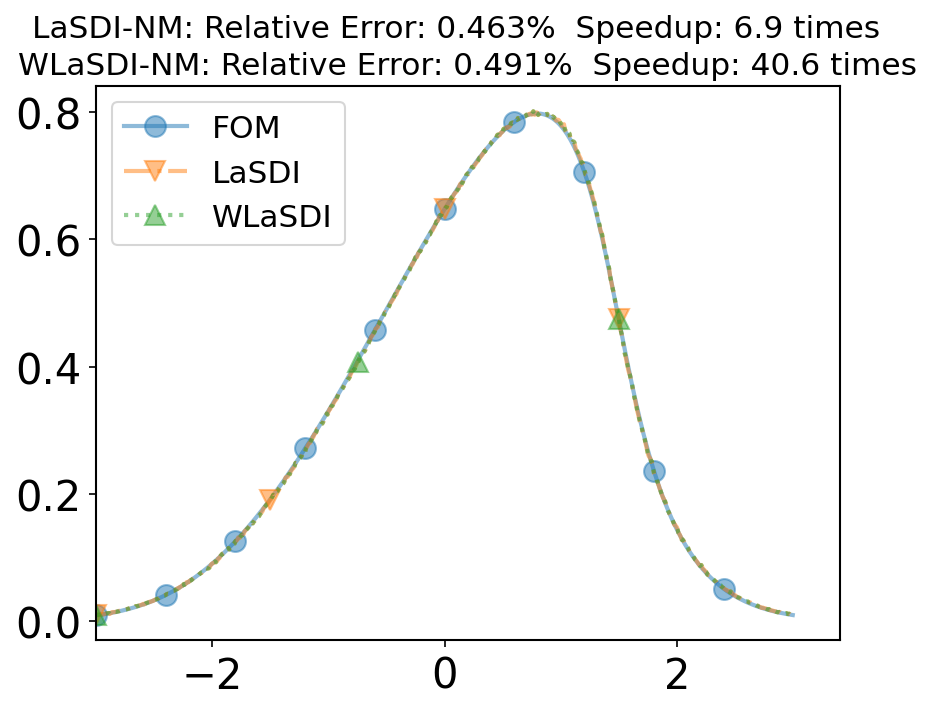}
        \caption{Using nonlinear compression}
    \end{subfigure}
    \caption{Recovered solutions of the noise-free 1D Burgers' equation at a = 0.8, w = 1.0 with 4 train points a = [0.85, 0.95] and w = [0.95, 1.05] at the final time step using
global WLaSDI and global LaSDI. The latent spaces’ dimensions are 5 and 4 respectively.}
    \label{fig:1DB_recover}
\end{figure}

Figure \ref{fig:1DB_recover} illustrates the final time step solution of the 1D Burgers' equation with no added noise. We conducted training with four data points, where $a$ took values in the range $[0.75, 0.85]$ and $w$ in the range $[0.95, 1.05]$, and for testing, we used $a = 0.8$ and $w = 1.0$. As observed in Figure \ref{fig:1DB_recover}, both the global WLaSDI and global LaSDI methods provide solutions that closely match the FOM solution. However, it is noteworthy that the WLaSDI approach offers higher speedups. Please note that speedup is determined by the ratio between the FOM simulation time and the time required to obtain the ROM solution 
In Figure \ref{fig:1DB_recover} (and in subsequent figures) note that the reported speedups are typical, but can change slightly depending on hardware and CPU load. We currently lack a complete understanding of the reasons behind the better speedup achieved by  WLaSDI compared to LaSDI in this case.

\begin{figure}
    \centering
    \begin{subfigure}[t]{0.7\textwidth} 
        \centering
        \includegraphics[width=\linewidth]{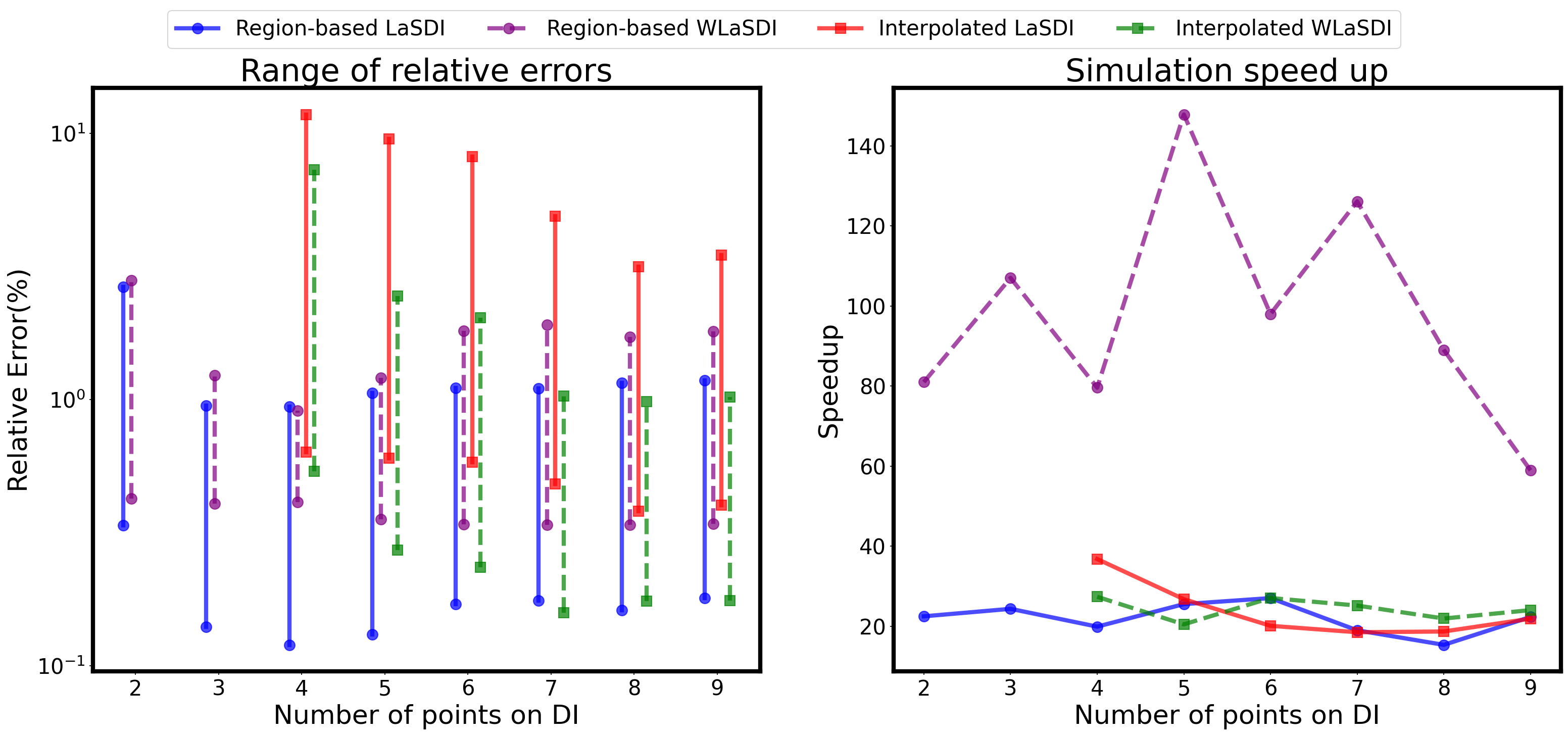}
        \caption{Using linear compression}
    \end{subfigure}
    \hfill
    \begin{subfigure}[t]{0.7\textwidth}
        \centering
        \includegraphics[width=\linewidth]{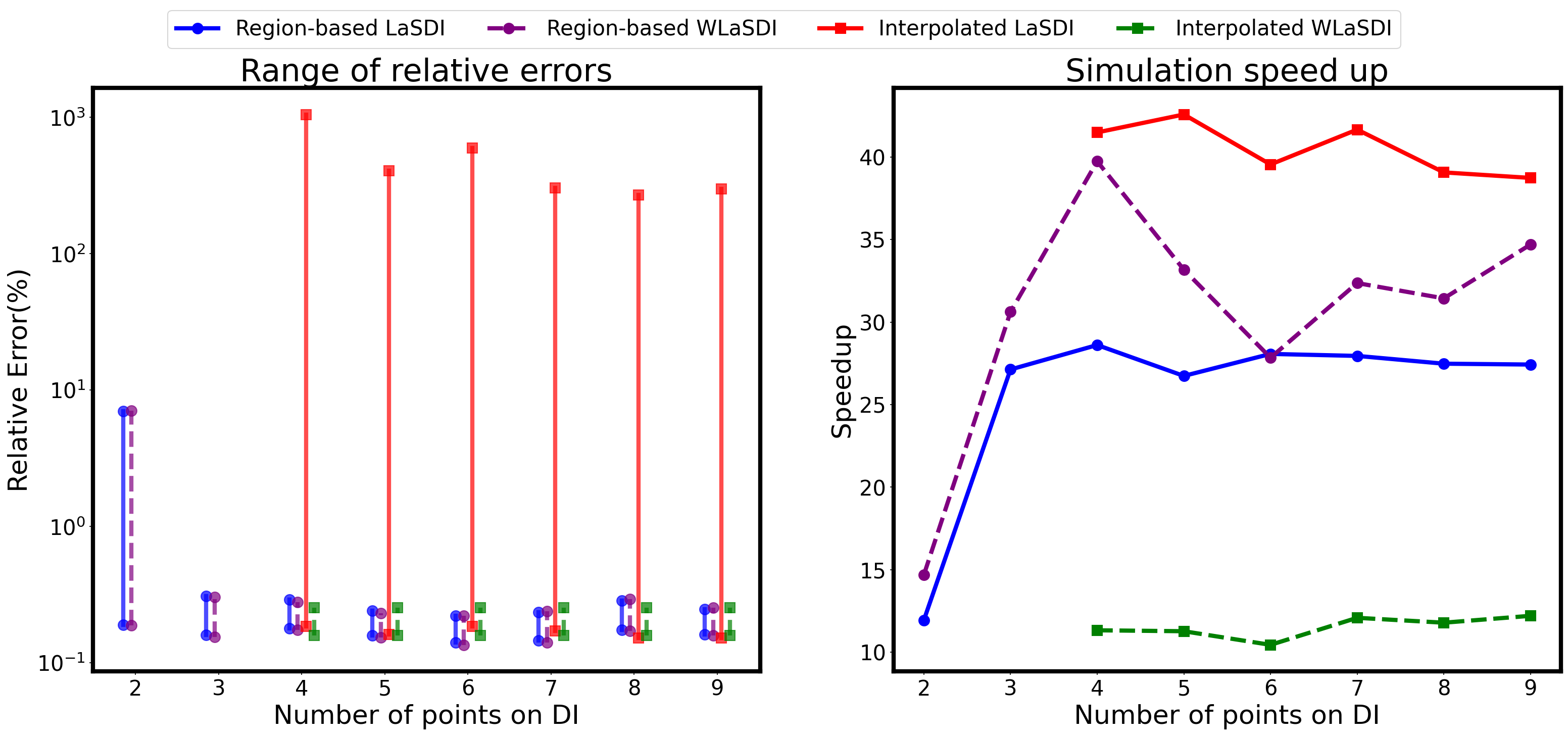}
        \caption{Using nonlinear compression}
    \end{subfigure}
   
    \caption{Performance Comparison: LaSDI vs. WLaSDI on the noise-free 1D Inviscid Burgers' Equation. The left-hand figures depict ranges in relative error as a function of the number of DI points, while the right-hand figures illustrate their corresponding speedups. Figure (a) showcases the results obtained with linear compression and figure (b) presents the outcomes using nonlinear compression. We use a consistent setup with 9 training points, a latent space dimension of 5 for LS methods, 4 for NM methods, and Degree 1 DI.} 
    \label{fig:1DB_ranges}
\end{figure}

To assess the performances of different local LaSDI and local WLaSDI without the presence of noise, we conducted a comparative analysis. Figure \ref{fig:1DB_ranges} illustrates the range of relative error and speedup for region-based and interpolated local methods and their speedup. We generated the full snapshot matrix using nine training points, $a = [0.7, 0.8, 0.9]$ and $w = [0.9, 1.0, 1.1]$. The size of the local DI, representing the number of points used for simulating the solution at $a = 0.85$ and $w = 0.95$, was systematically varied from 2 to 9. 
According to findings in \cite{FriesHeChoi2022ComputerMethodsinAppliedMechanicsandEngineering}, for the 1D Burgers' equation, linear compression outperforms nonlinear compression. In Figure \ref{fig:1DB_ranges}a, 
it is evident that region-based LaSDI outperforms WLaSDI, albeit by a small margin. This difference in performance may be attributed to potential additional computational errors associated with the weak form. The speedup is comparable between interpolated WLaSDI and LaSDI, but a significant improvement is observed when using the region-based, as depicted in the top-right figure of Figure \ref{fig:1DB_ranges}. However, at this point, the exact reasons for this speedup in this particular case remain uncertain.
When nonlinear compression is employed, while the region-based models between LaSDI and WLaSDI show comparable results, interpolated WLaSDI outperforms interpolated LaSDI. This could be attributed to the possibility that LaSDI is returning unstable governing equations for the latent space or is using  an unsuitable interpolation method.  

\begin{figure}
\centering
\begin{subfigure}[t]{0.45\textwidth}
        \centering
        \includegraphics[width=\linewidth]{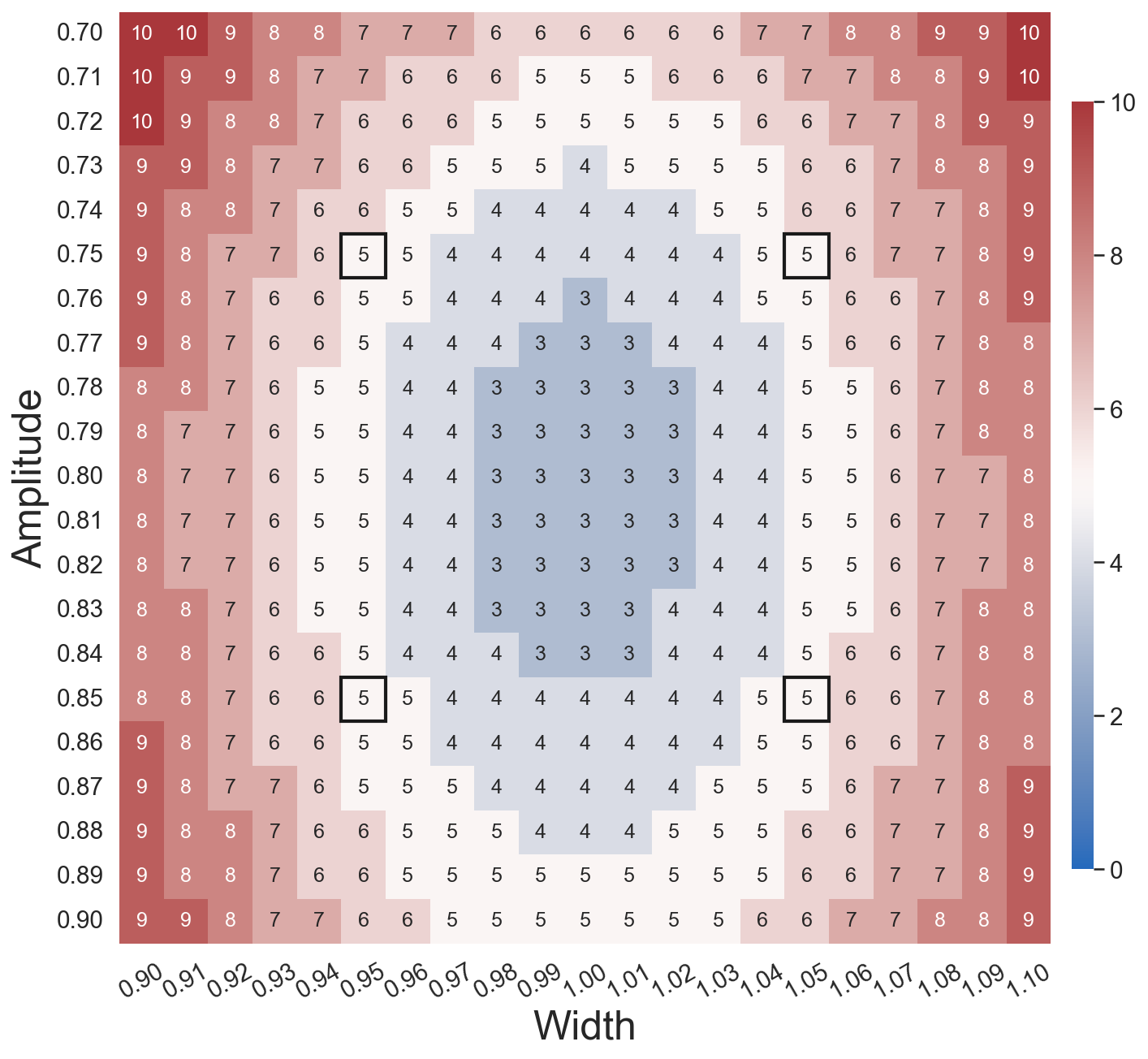}
        \caption{Max relative error of LaSDI - LS}
    \end{subfigure}
    \begin{subfigure}[t]{0.45\textwidth}
        \centering
        \includegraphics[width=\linewidth]{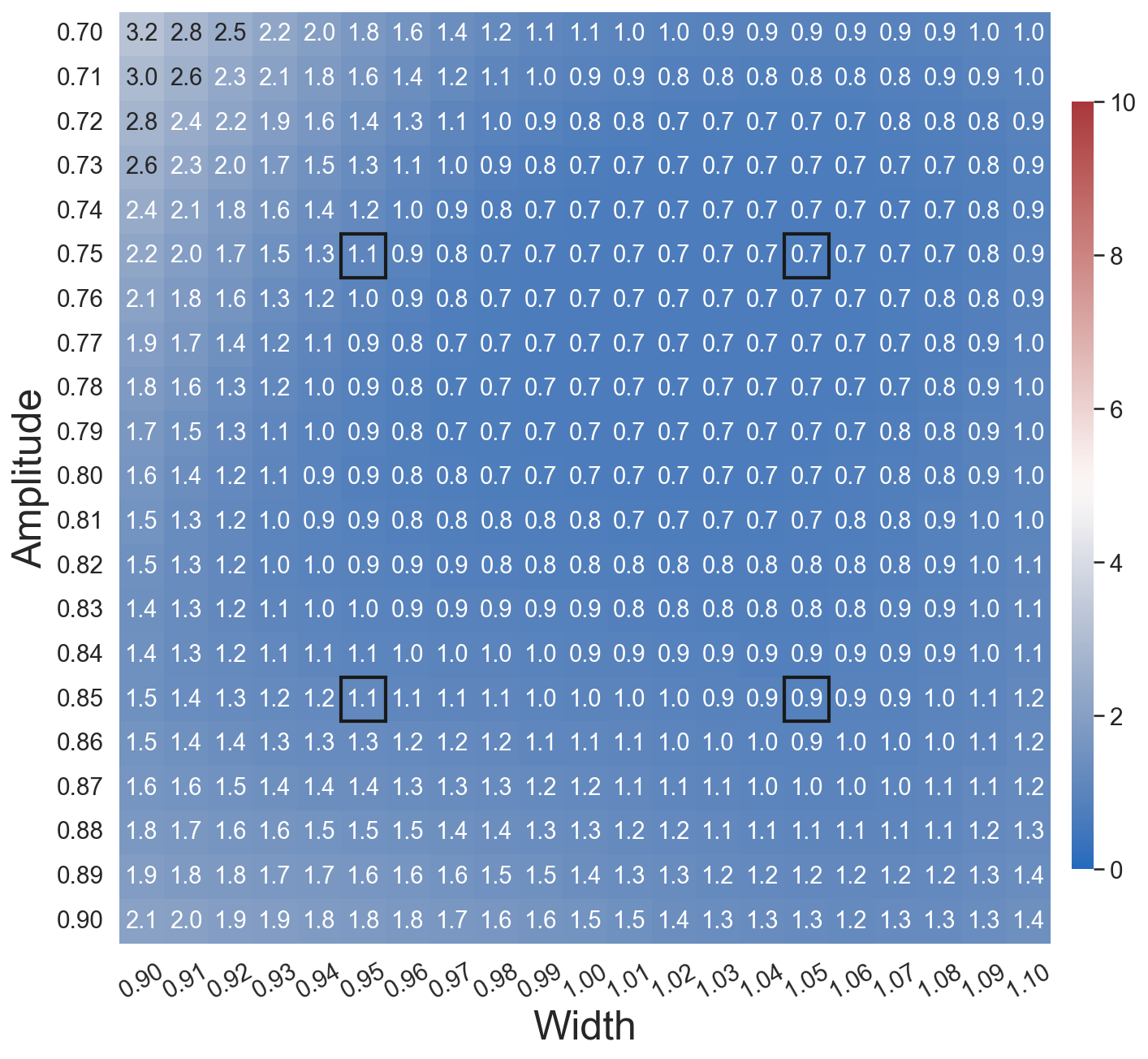}
        \caption{Max relative error of WLaSDI - LS}
    \end{subfigure}
    \hfill
    \begin{subfigure}[t]{0.45\textwidth}
        \centering
        \includegraphics[width=\linewidth]{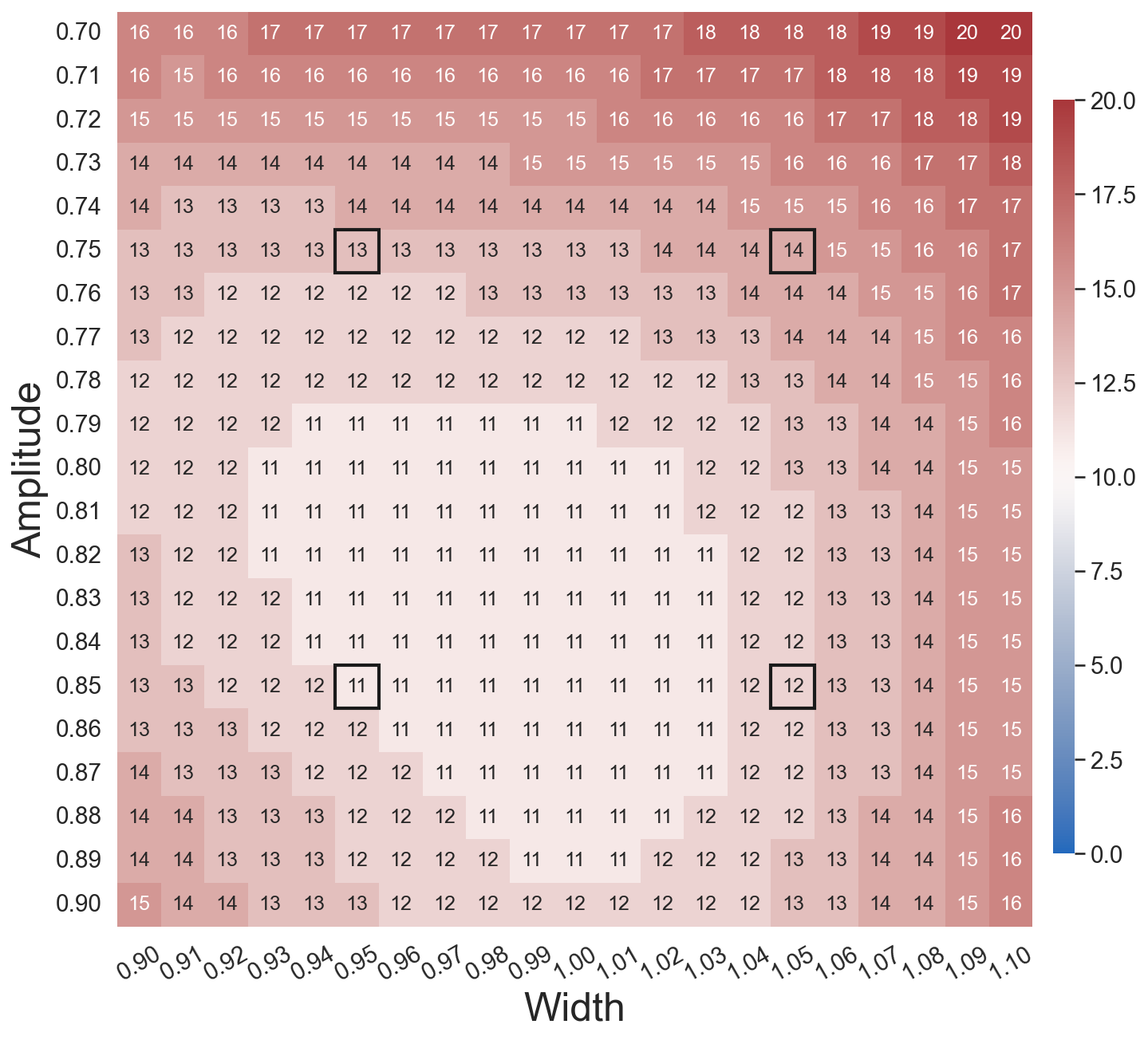}
        \caption{Max relative error of LaSDI - NM}
    \end{subfigure}
    \begin{subfigure}[t]{0.45\textwidth}
        \centering
        \includegraphics[width=\linewidth]{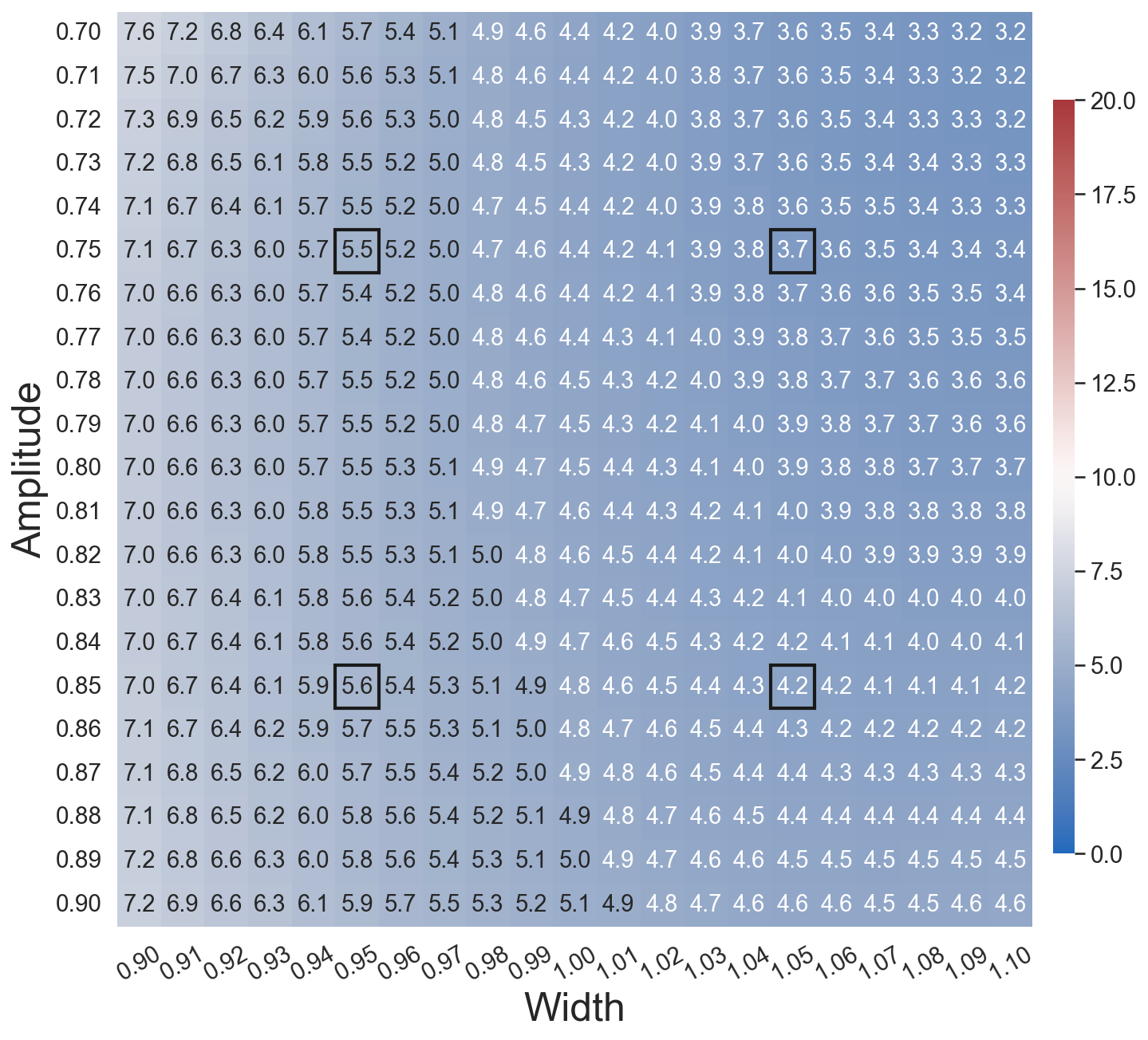}
        \caption{Max relative error of WLaSDI - NM}
    \end{subfigure}
    \caption{Mean maximum relative error of the 1D Burgers' equation on the whole parameter space with 4 training points $a = [0.75, 0.85]$ and $w = [0.95, 1,05]$. Utilizing global LaSDI and global WLaSDI, along with both compression methods, we employ latent space dimensions of 5 and 4, respectively. Degree 1 dynamical systems are consistently applied. Training points in the parameter space are marked with black squares. $10 \%$ white noise is added to the training data. }
    \label{fig:1DB_heatmap}
\end{figure}

We introduce a 10\% Gaussian white noise component to the 1D Burgers' measurement data and assess the performance of LaSDI and WLaSDI. In Figure \ref{fig:1DB_heatmap}, we present heat maps illustrating the mean maximum relative error, which is given by the average of 
$E(\mathbf{U)} = \max_{t} \frac{\left\Vert \mathbf{U}(t) -  \widetilde{\mathbf{U}}(t)\right\Vert_2}{\left\Vert\mathbf{U}(t)\right\Vert_2}$ where $\widetilde{\mathbf{U}}(t)$ is the recovered FOM from WLaSDI.  
The mean maximum relative error is taken across the entire parameter space $\mathcal{D}$ when using global LaSDI and WLaSDI. It can clearly be seen that the global WLaSDI exhibits significantly greater robustness to noise. With a 10\% noise level, WLaSDI produces a model with less than a 4\% relative error using linear compression, whereas LaSDI records a 10\% error. Similarly, the errors for WLaSDI-NM and LaSDI-NM are 8\% and 20\%, respectively, as shown in Figure \ref{fig:1DB_heatmap}(c-d).

\begin{figure}[h]
  \centering
  {\includegraphics[width=0.7\textwidth]{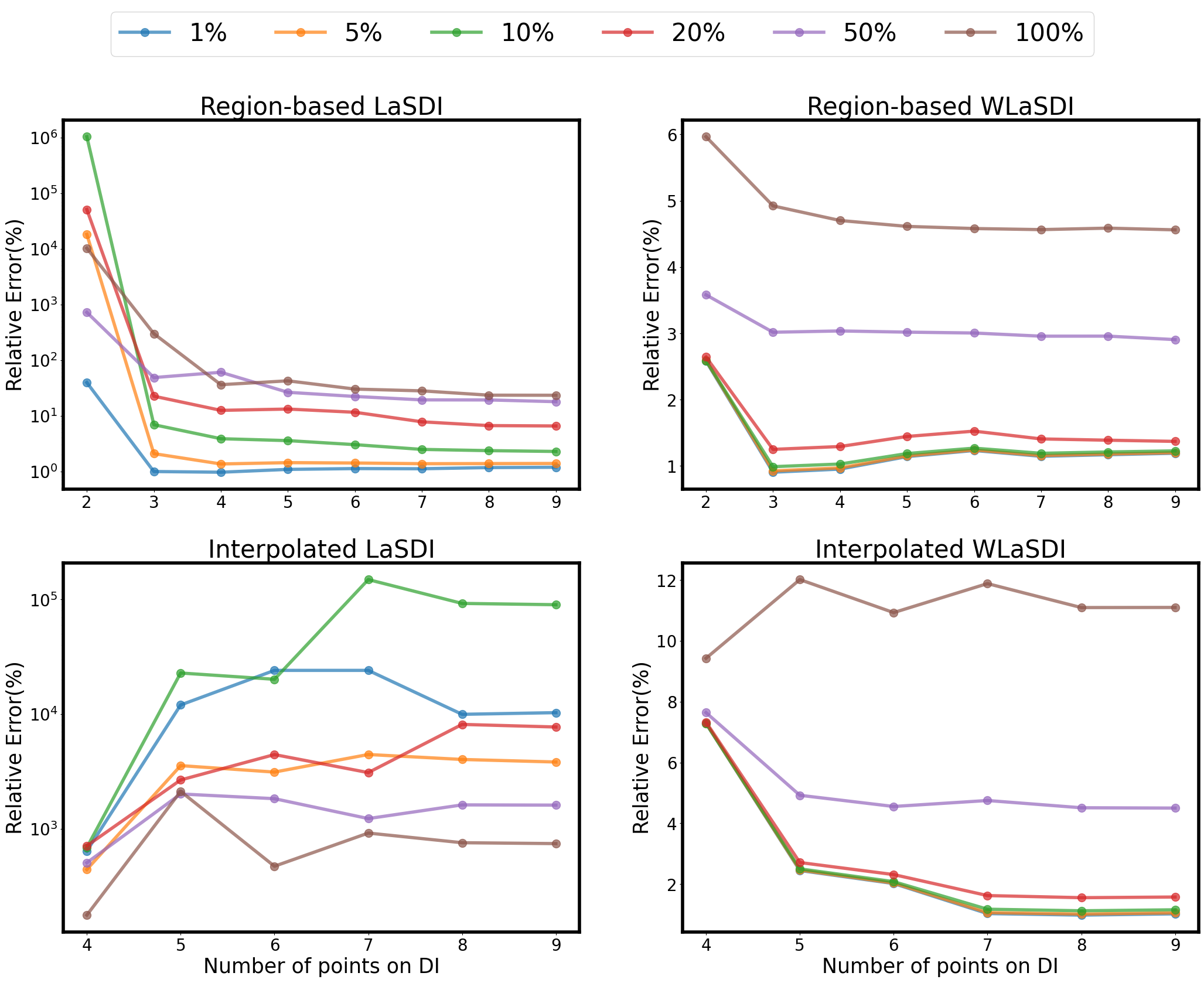}}
  \hfill
   \caption{Mean maximum relative errors of the 1D Burgers' equation across different local techniques as we vary the number of training points used in DI and noise levels. Utilizing 9 training points ($a = [0.7, 0.8, 0.9]$ and $w = [0.9, 1.0, 1.1]$), we focus on a testing point at $a = 0.85$ and $w = 0.95$. The analysis encompasses linear compression and latent space dimensions of 5, with Degree 1 DI.}
   \label{fig:1DB_LS_noise}
\end{figure}

\begin{figure}[h]
  \centering
  {\includegraphics[width=0.7\textwidth]{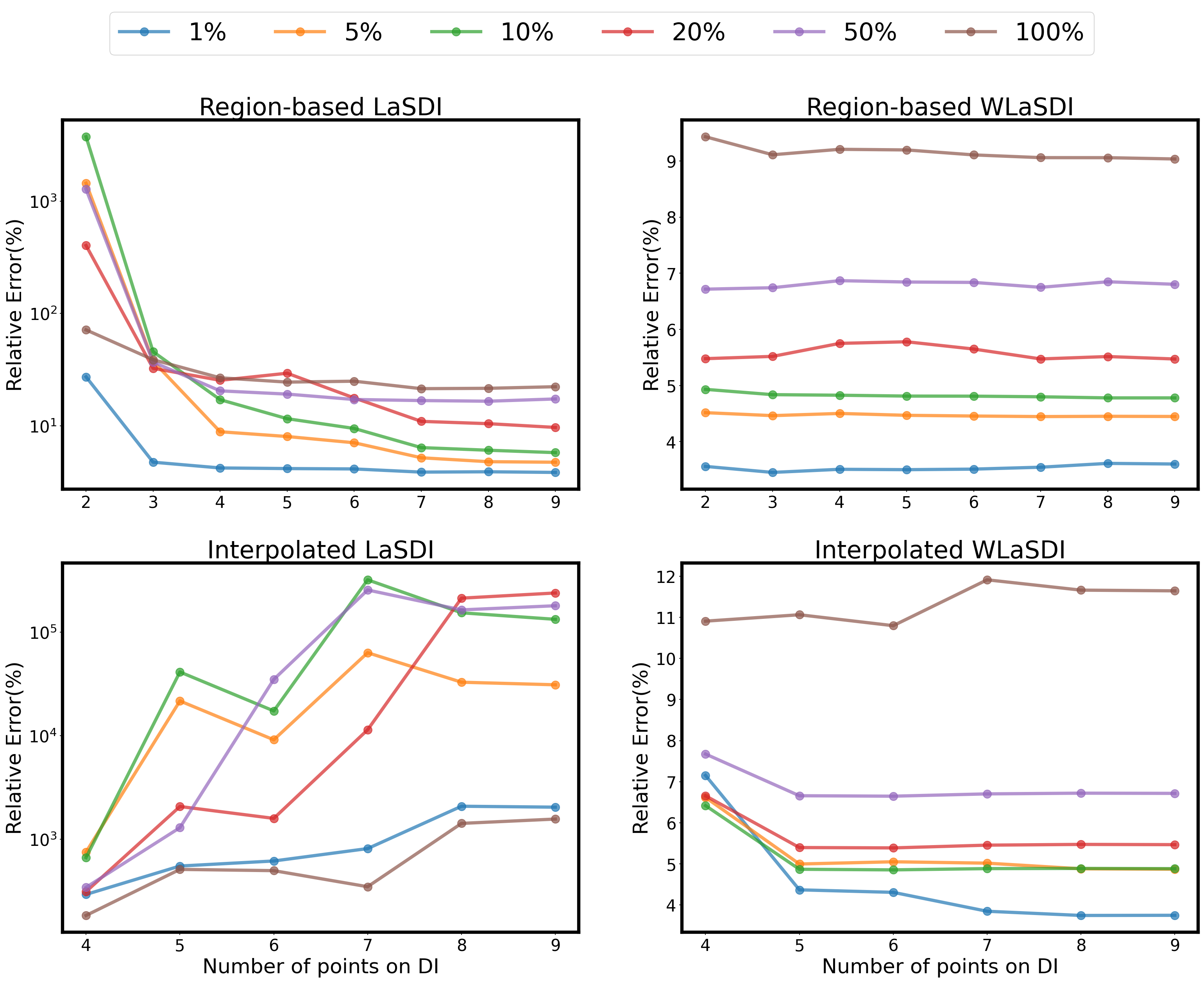}}
   \caption{Mean maximum relative errors of the 1D Burgers' equation across different local techniques as we vary the number of training points used in DI and noise levels. Utilizing 9 training points ($a = [0.7, 0.8, 0.9]$ and $w = [0.9, 1.0, 1.1]$), we focus on a testing point at $a = 0.85$ and $w = 0.95$. The analysis encompasses nonlinear compression and latent space dimensions of 4, with Degree 1 DI.}
   \label{fig:1DB_NM_noise}
\end{figure}

In our final analysis for 1D Burgers', as illustrated in Figure \ref{fig:1DB_LS_noise} and \ref{fig:1DB_NM_noise}, we delve into the performance of WLaSDI under various noise levels, including 1\%, 5\%, 10\%, 20\%, 50\%, and 100\%. This investigation distinctly demonstrates the remarkable robustness of WLaSDI when compared to LaSDI. In fact, when subjected to a substantial 100\% Gaussian noise, region-based WLaSDI yields maximum relative errors of less than 6\% and 12\% for linear and nonlinear compression, respectively. On the other hand, LaSDI returns significantly higher figures, with maximum relative errors reaching up to $10,000\%$ with the interpolated case. This could be due to several reasons. It is possible that the ODEs returned by LaSDI are unstable, which may require fine-tuning the regularization parameters when running SINDy. Furthermore, this issue may also be related to the interpolation method in use. While we currently use radial basis function interpolation, it is worth considering alternative methods that might yield better results. Proper tuning of the interpolation parameters is also essential. 


This analysis strongly underscores the superior performance of WLaSDI, particularly in scenarios involving high noise levels. Additionally, it is worth noting that the mean maximum relative error values presented in this section are computed by averaging the maximum relative errors across $100$ instances, ensuring statistical significance.

\subsection{2D Viscous Burgers'}\label{sec:2DB}
\begin{figure}[h]
    \centering
   \begin{subfigure}[t]{0.7\textwidth}
   \centering
    {\includegraphics[width=\linewidth]{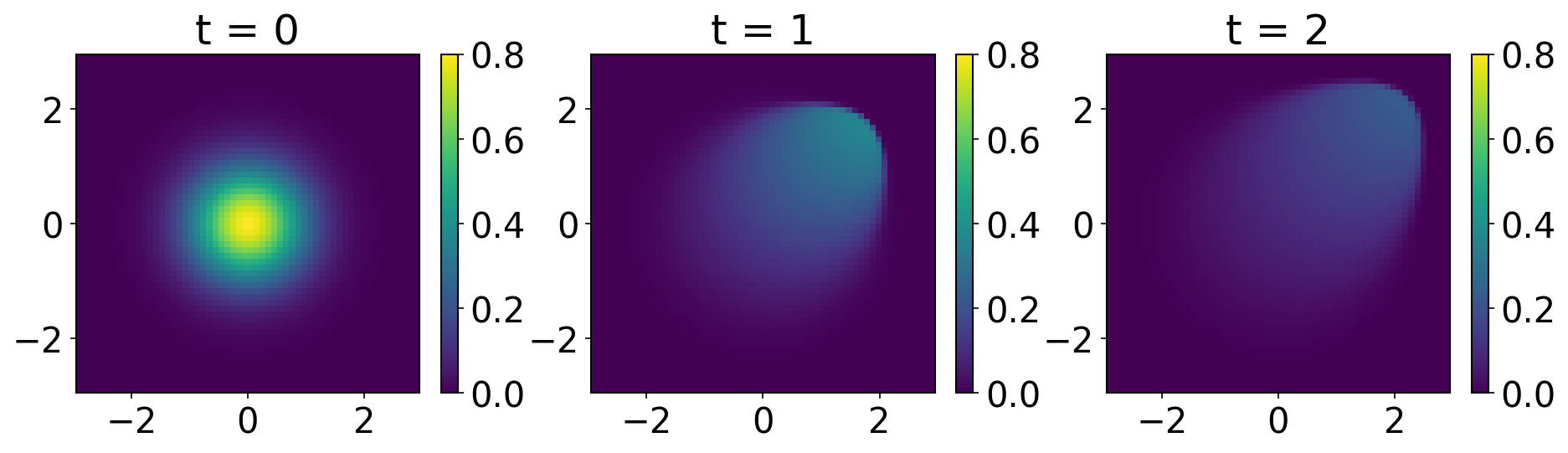}}
    \caption{Numerical solution to the 2D Burgers' equation}
   \end{subfigure}
  \begin{subfigure}[t]{0.30\textwidth}
        \centering
        \includegraphics[width=\linewidth]{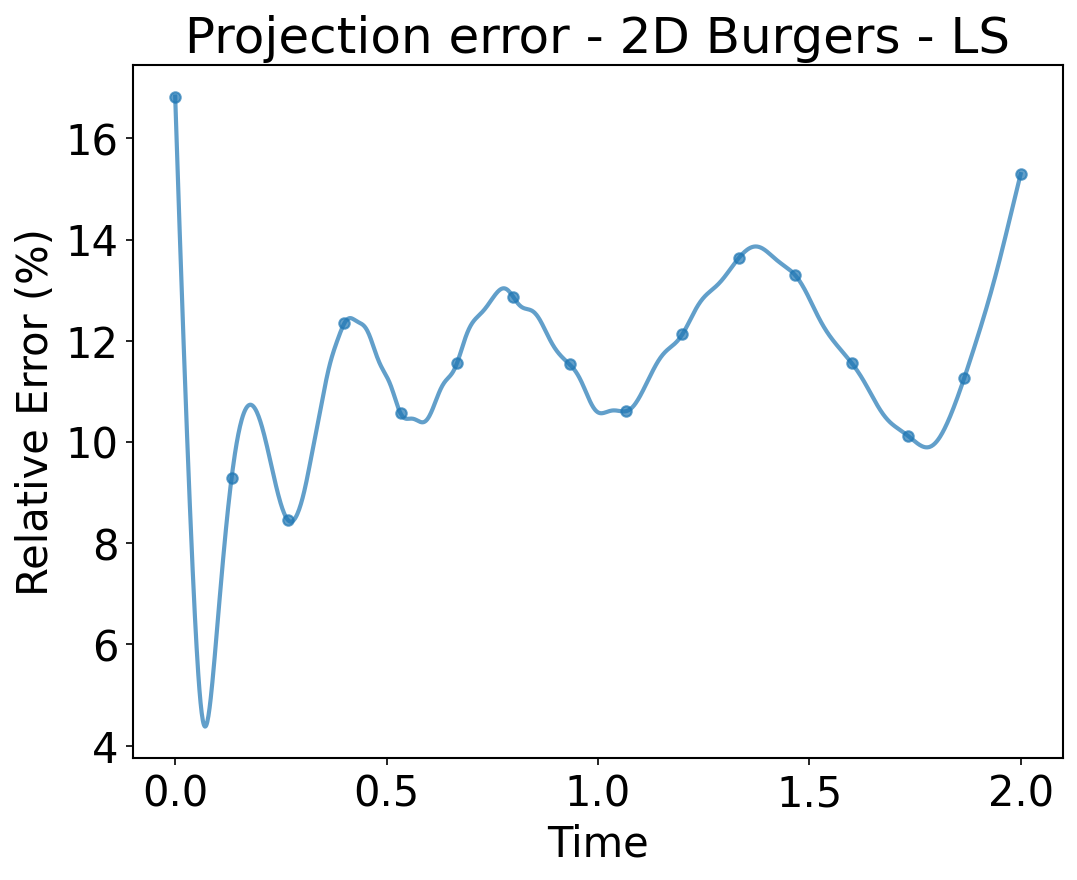}
        \caption{}
 \end{subfigure}
 \begin{subfigure}[t]{0.3\textwidth}
        \centering
        \includegraphics[width=\linewidth]{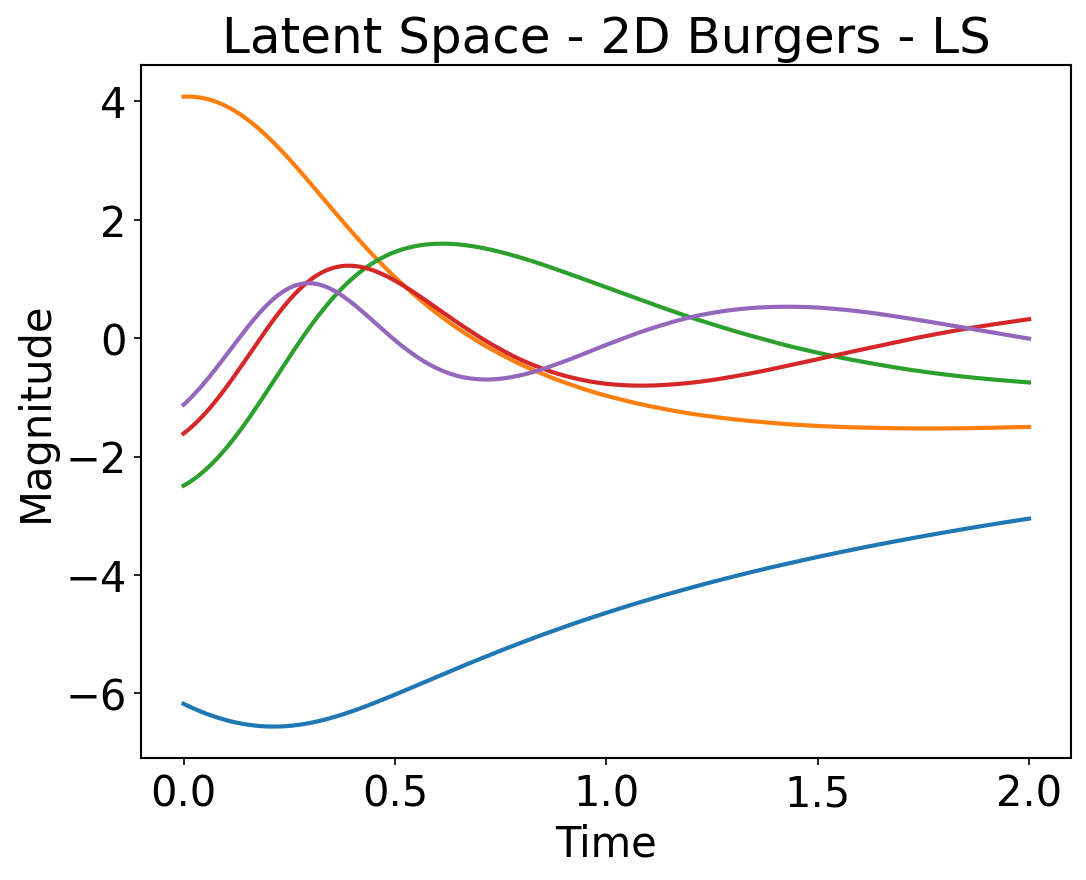}
        \caption{}
 \end{subfigure}
  \hfill
   \caption{The 2D Burgers' equation. Figure (a) is the solution to the 2D viscous Burger's equation. Figure (b) is the projection error, which highlights the disparity between the original data and the compressed - decompressed representation. Figure (c) is the latent space trajectory obtained through linear compression.}
   \label{fig:2DB_intro}
\end{figure}

For the 2D  Viscous Burgers' equation, Figure \ref{fig:2DB_intro} showcases an example solution to the 2D Burgers' equation, as well as the projection error resulting from linear compression and the corresponding latent space trajectories. We chose a latent space dimension of 5 based on our analysis of the cumulative sum of singular values, where we performed truncation to retain 99\% of the total energy. The projection error represents the relative difference between the true data and the decompressed-compressed representation. We utilize a dataset with nine training points and exclusively employ proper orthogonal decomposition for compression. Our primary objective is to assess the performance of WLaSDI and LaSDI under varying noise levels.

\begin{figure}[h]
  \centering
  {\includegraphics[width=0.7\textwidth]{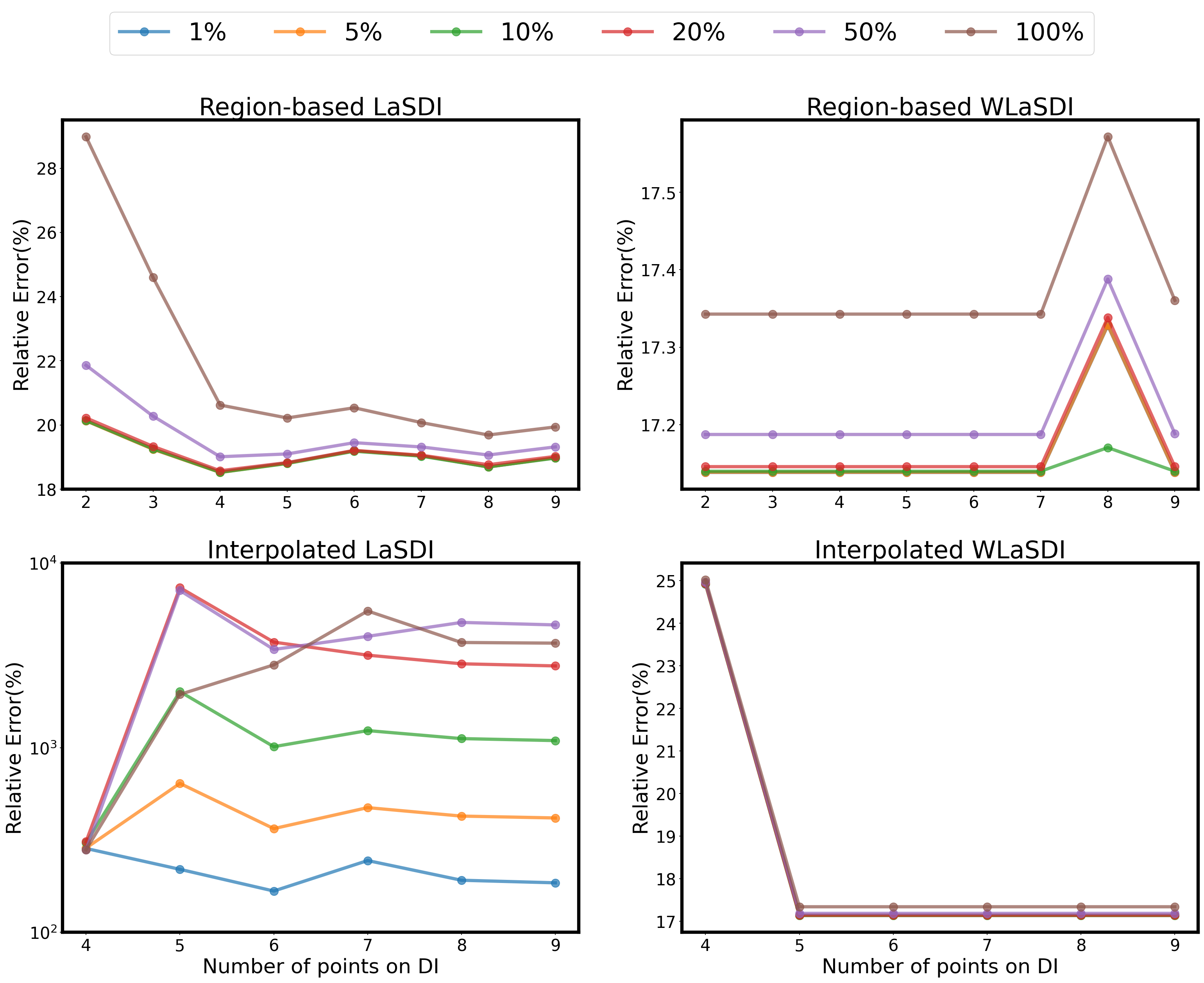}}
   \caption{Mean maximum relative errors for the 2D viscous Burgers' equation using various local techniques. We vary the number of training points and noise levels. We use nine training points ($a = [0.7, 0.8, 0.9]$ and $w = [0.9, 1.0, 1.1]$) and focus on a testing point at $a = 0.85$ and $w = 0.95$. The analysis includes linear compression and a latent space dimension of 5, with Degree 1 DI.} 
   \label{fig:2DB_LS_noise}
\end{figure}
Upon a close examination of Figure \ref{fig:2DB_LS_noise}, a distinct trend emerges – the Weak form is the consistent performance winner. When employing a region-based approach, the mean maximum relative error of LaSDI on the 2D Viscous Burgers' equation fluctuates within the range of 18\% to 30\% across various noise levels. In stark contrast, the same metric remains below 18\% for WLaSDI. Considering the projection error in Figure \ref{fig:2DB_intro}, where the projection error at $t = 0$ is $16\%$, the mean maximum relative error using WLaSDI is not expected to fall below 18\%. Taking a closer look at the interpolated method, the disparity in performance becomes even more pronounced. For LaSDI, the mean maximum relative error experiences a considerable increase across all noise levels, whereas it remains below 25\% for WLaSDI. It may appear counterintuitive that LaSDI performs better with higher noise levels than with lower ones, but all of these results pale in comparison to the performance of WLaSDI. This discrepancy underscores the robustness of WLaSDI in the face of noise, clearly demonstrating its superiority over LaSDI. It is important to note that these observations are drawn from an extensive analysis involving 100 instances. 

\subsection{Heat Conduction}\label{sec:Diff}
\begin{figure}[h]
    \centering
   \begin{subfigure}[t]{0.7\textwidth}
   \centering
    {\includegraphics[width=\linewidth]{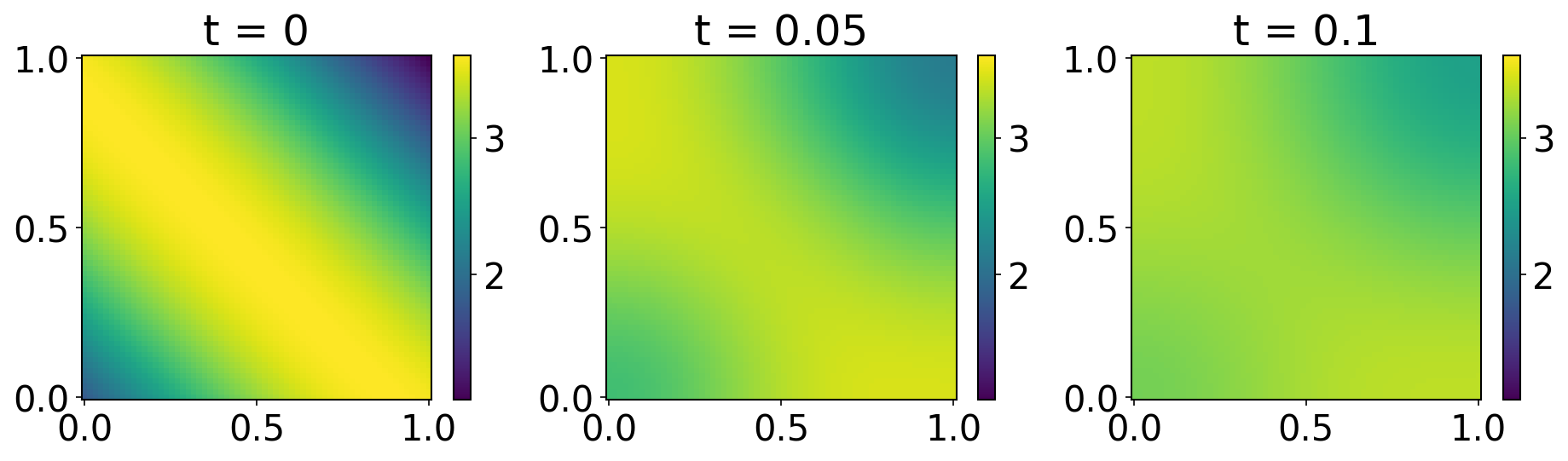}}
    \caption{Numerical solution to the Heat Conduction equation}
   \end{subfigure}
  \begin{subfigure}[t]{0.30\textwidth}
        \centering
        \includegraphics[width=\linewidth]{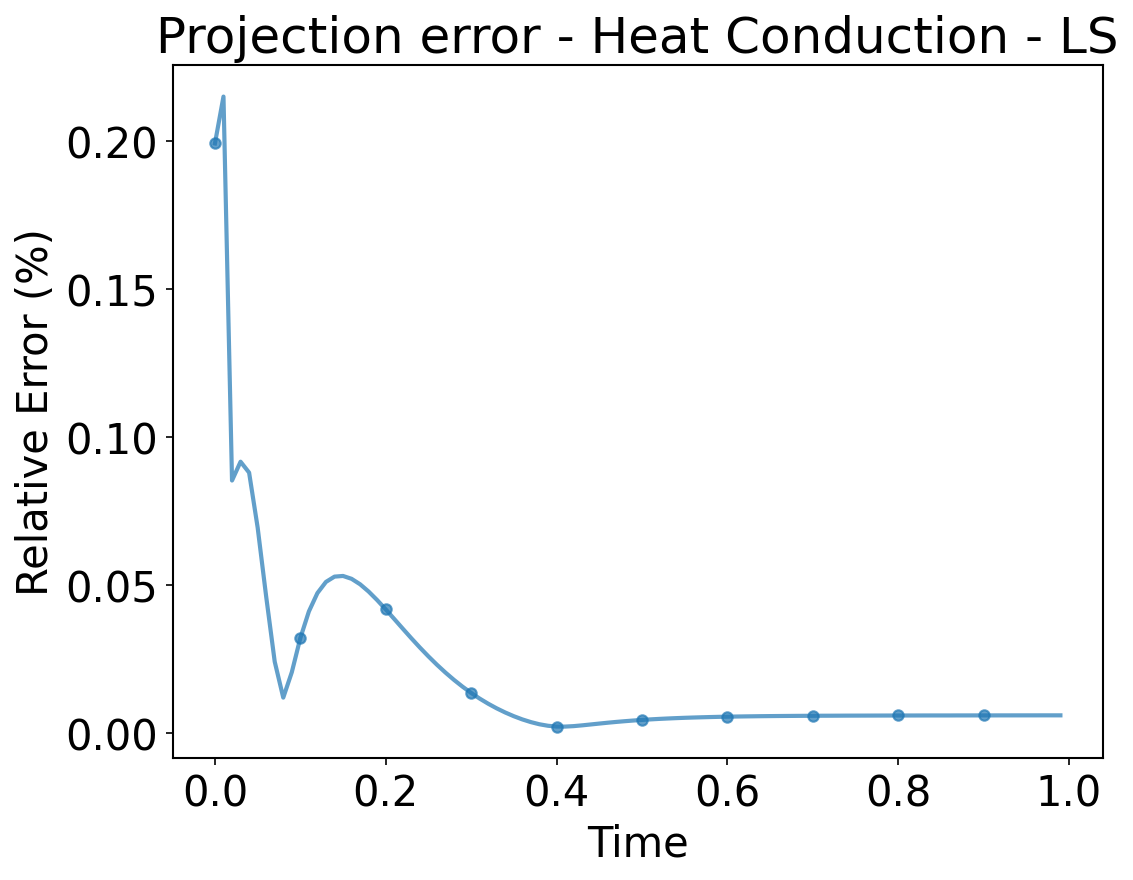}
        \caption{}
 \end{subfigure}
 \begin{subfigure}[t]{0.3\textwidth}
        \centering
        \includegraphics[width=\linewidth]{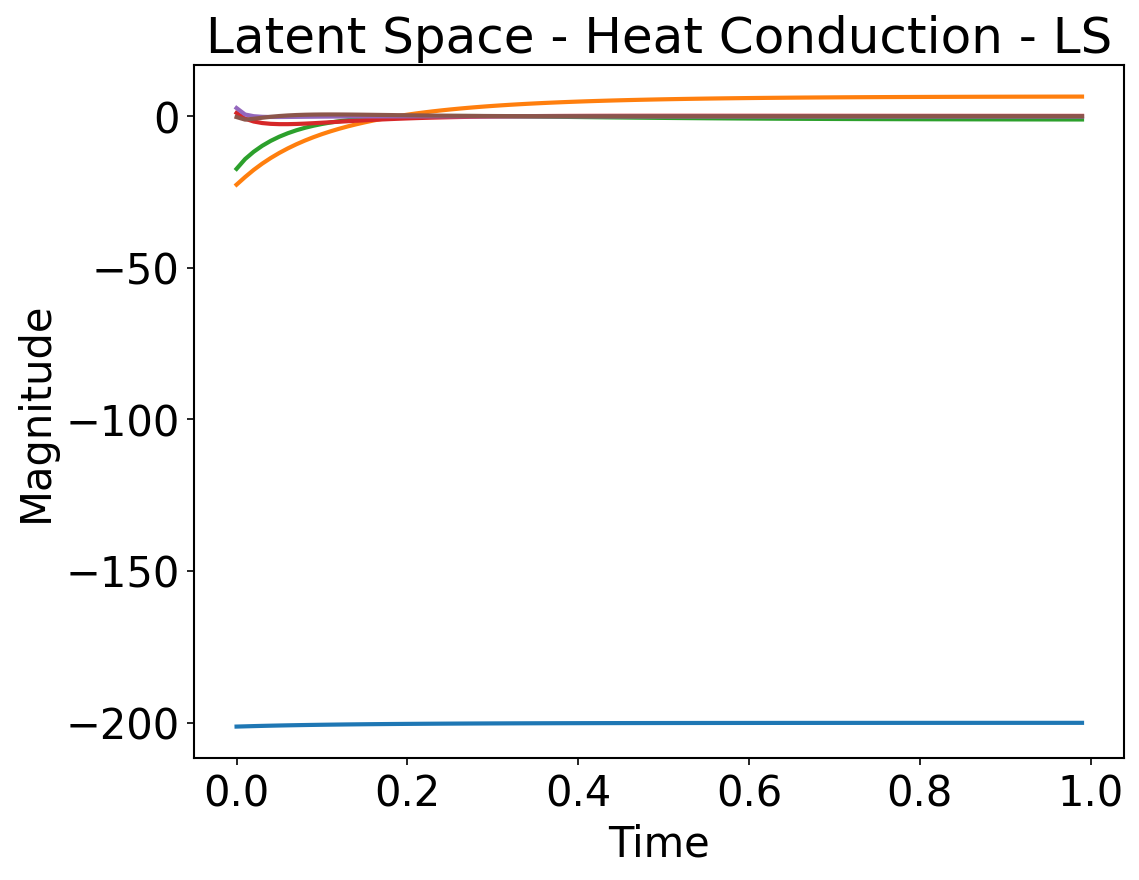}
        \caption{}
 \end{subfigure}
  \hfill
   \caption{The Heat Conduction equation. Figure a is the solution to the heat conduction equation. Figure b is the projection error, which highlights the disparity between the original data and the compressed representation. Figure c is the latent space trajectory obtained through linear compression.}
   \label{fig:diff_intro}
\end{figure}

For the heat conduction equation, Figure \ref{fig:diff_intro}  illustrates a solution, alongside the projection error resulting from linear compression and the corresponding trajectories in the latent space. We have selected a latent space dimension of 6 after evaluating the cumulative sum of singular values and applying truncation to retain 99\% of the energy. In this particular example, we utilize 9 training data points and apply the linear compression technique to explore the capabilities and robustness of WLaSDI when dealing with the intricacies of the heat conduction equation.

\begin{figure}[h]
  \centering
  {\includegraphics[width=0.7\textwidth]{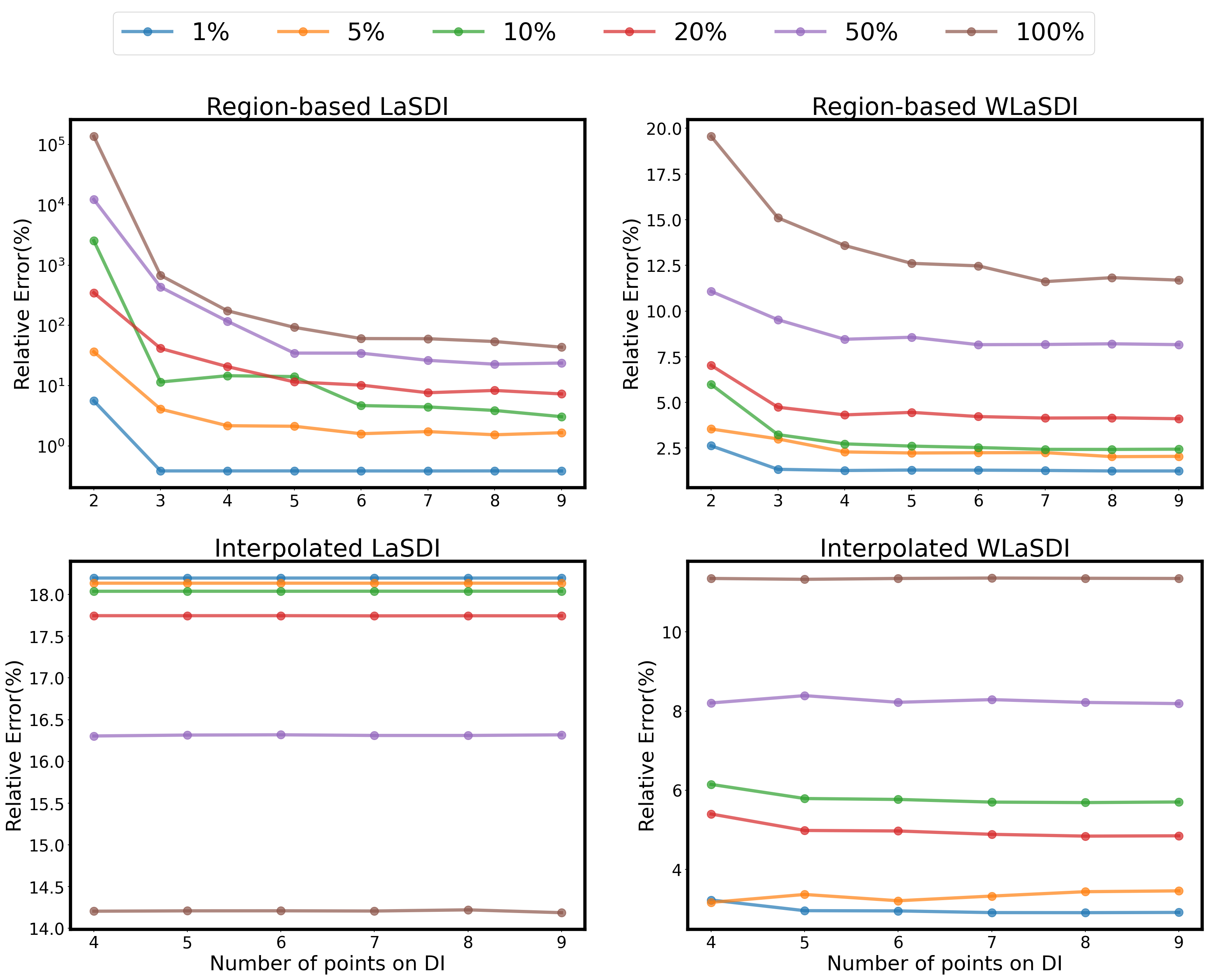}}
   \caption{Mean maximum relative errors for the Heat Conduction equation for various local techniques as we change the number of points used in the dynamics identification step under different noise levels. We employed 9 training points and a latent space dimension of 6 while utilizing Degree 1 DI. The analysis includes linear compression and a latent space dimension of 5, with Degree 1 DI.}
   \label{fig:diff_LS_local}
\end{figure}

It is clear that region-based WLaSDI outperforms region-based LaSDI, as demonstrated in Figure \ref{fig:diff_LS_local}, especially in scenarios characterized by high levels of noise. When subjected to 100\% noise, WLaSDI displays a mean maximum relative error of less than 20\%, while LaSDI's mean maximum error exceeds 100\%. Moreover, even when exposed to 100\% noise, interpolated WLaSDI's results remain below 11\%, in contrast to LaSDI, where the error rate is higher. It is important to note that the mean maximum relative error is determined by averaging the maximum relative errors across 100 instances.

 \subsection{Radial Advection}\label{sec:RA}
\begin{figure}[h]
    \centering
   \begin{subfigure}[t]{0.7\textwidth}
   \centering
    {\includegraphics[width=\linewidth]{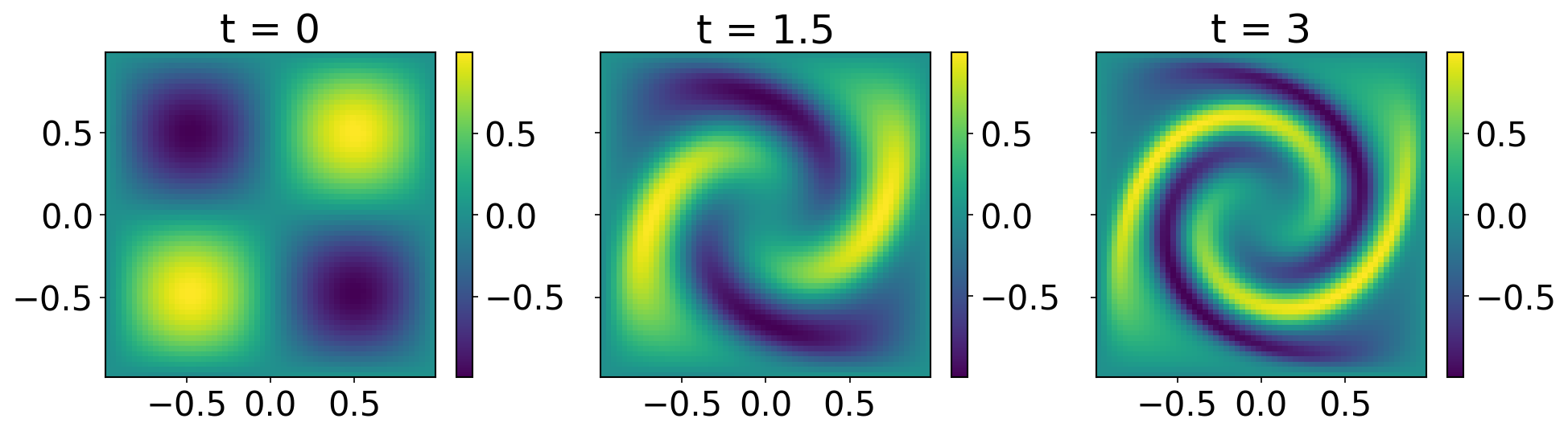}}
    \caption{Numerical solution to the Radial Advection equation}
   \end{subfigure}
  \begin{subfigure}[t]{0.30\textwidth}
        \centering
        \includegraphics[width=\linewidth]{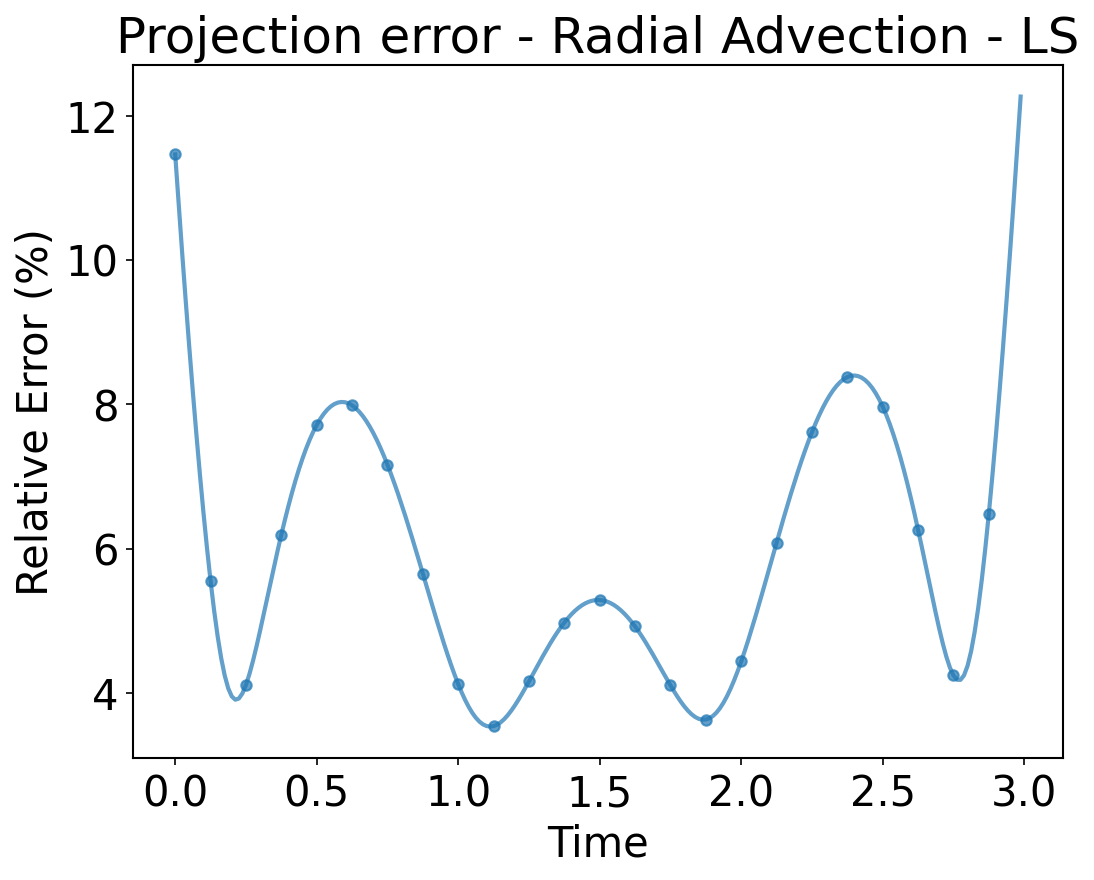}
        \caption{}
 \end{subfigure}
 \begin{subfigure}[t]{0.3\textwidth}
        \centering
        \includegraphics[width=\linewidth]{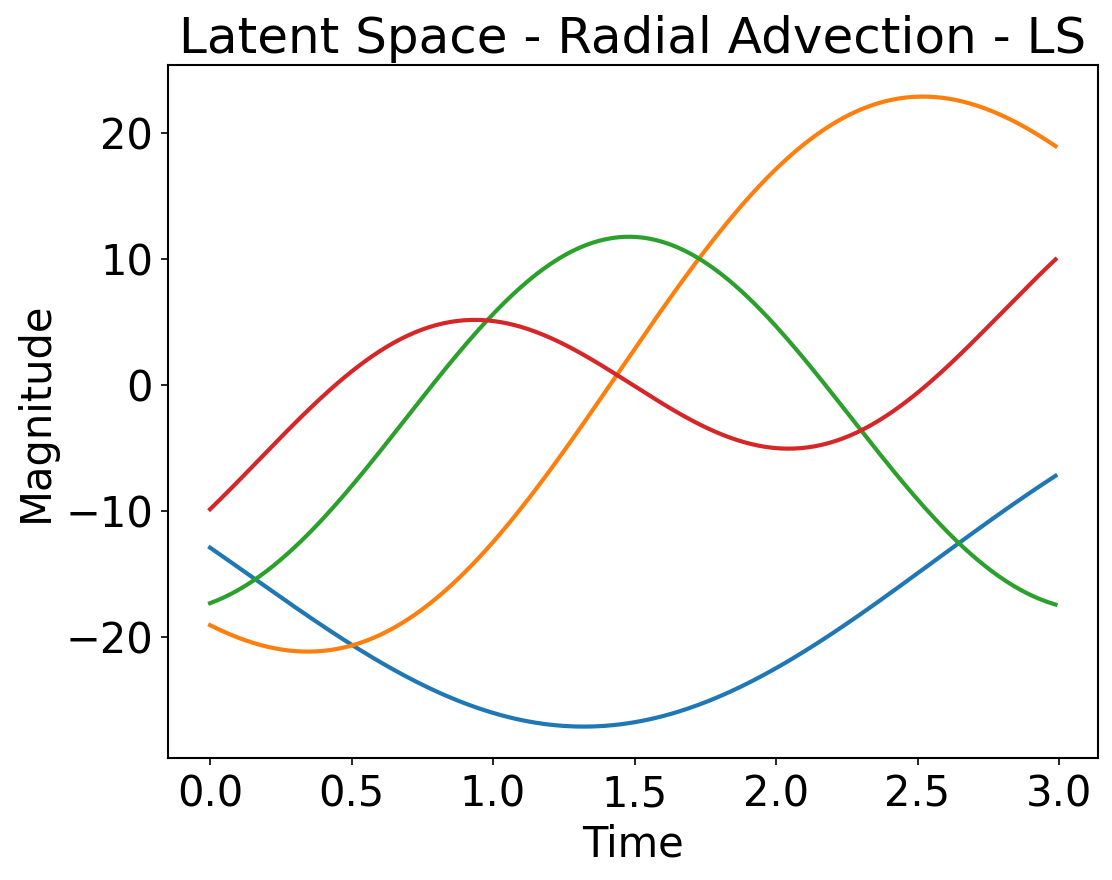}
        \caption{}
 \end{subfigure}
  \hfill
   \caption{The Radial Advection equation. Figure a is the solution to the radial advection equation. Figure b is the projection error, which highlights the disparity between the original data and the compressed representation. Figure c is the latent space trajectory obtained through linear compression.}
   \label{fig:RA_intro}
\end{figure}

 For the radial advection problem, Figure \ref{fig:RA_intro} illustrates a solution and the projection error resulting from linear compression and the associated trajectories in the latent space. We select a latent space dimension of 4 because, in this case, we assess the cumulative sum of singular values and trim it to retain 90\% of the energy. As in the previous section, we apply linear compression techniques and use a training dataset consisting of 9 points.

\begin{figure}[h]
  \centering
  {\includegraphics[width=0.7\textwidth]{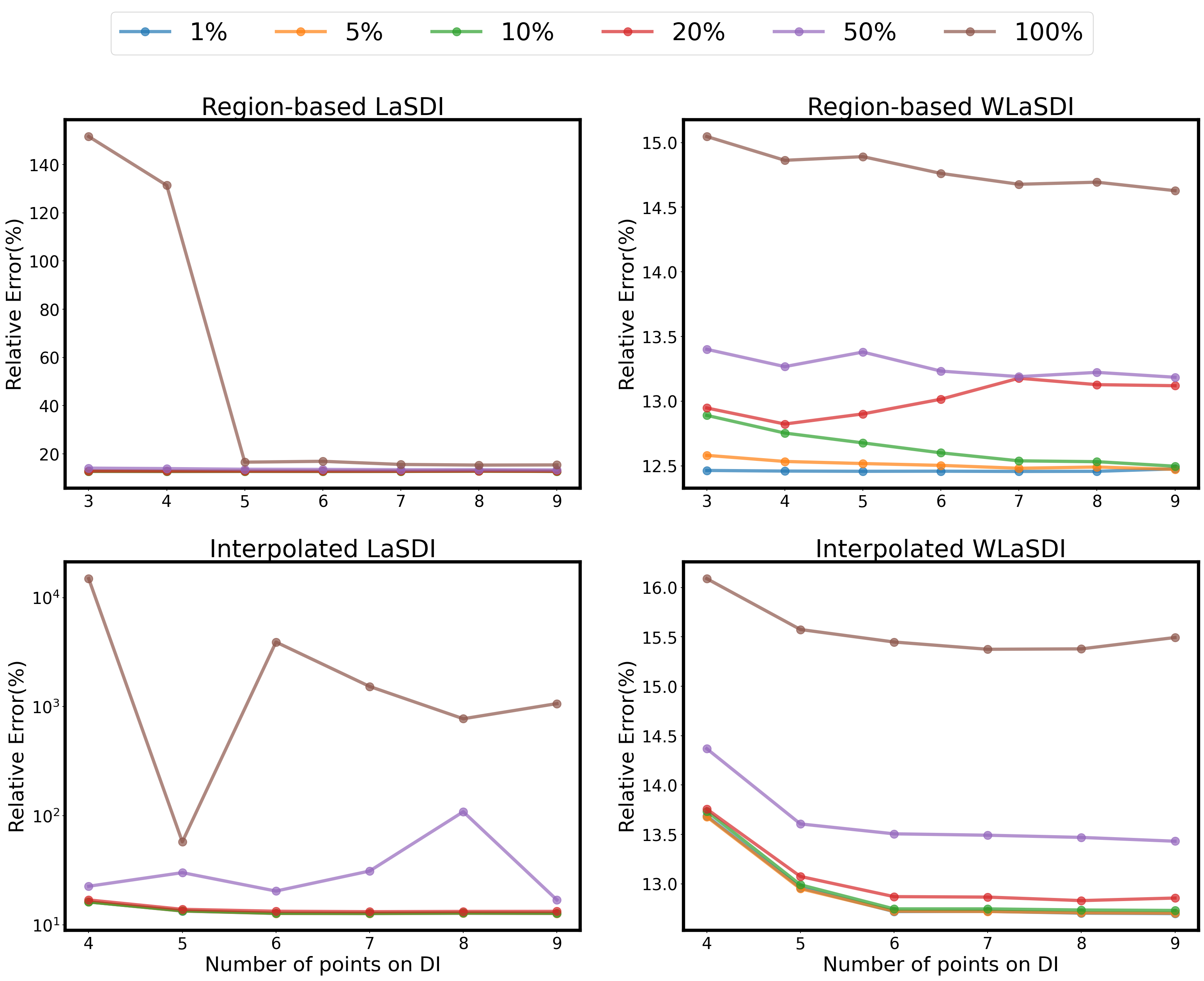}}
   \caption{Mean maximum relative errors of the Radial Advection equation for local techniques as we change the number of points used in the dynamics identification step under different noise levels. We employed 9 training points and a latent space dimension of 4 while utilizing degree 1 DI for both WLaSDI and LaSDI.}
   \label{fig:RA_LS_local}
\end{figure}

We opted for Degree 1 DI for both WLaSDI and LaSDI I. When assessing the comparison between region-based WLaSDI and LaSDI, it becomes evident that WLaSDI delivers slightly better results compared to LaSDI, as indicated in Figure \ref{fig:RA_LS_local}. Indeed, as the noise level increases, the performance gap in favor of WLaSDI becomes more pronounced. At 100\% noise, the relative error is under 16\% for WLaSDI, while it is much higher for LaSDI. Furthermore, in this example, similar results can be found with Degree 2 DI for WLaSDI, whereas for LaSDI, Degree 2 DI is unstable.

\section{Conclusion}\label{sec:conclusion}

We have introduced a new framework, WLaSDI: Weak form Latent Space Dynamics Identification that constructs robust and efficient parametric models. The main message of our work is that the inclusion of the weak form into the LaSDI framework provides substantial robustness to noise.

We used two compression techniques (POD and an autoencoder) and two distinct dynamics identification approaches (region-based and interpolated). Notably, we leverage weak form parameter estimation WENDy framework \cite{BortzMessengerDukic2023BullMathBiol} in the dynamics identification step to enhance robustness, particularly in the presence of noise. We also showed how our framework is applicable to a wide spectrum of physical phenomena, including viscous and inviscid Burgers' equations, radial advection, and heat conduction. The utilization of weak form parameter estimation offers several advantages over the previously employed strong form in LaSDI. It clearly yields superior results in the presence of noise. The weak form also provides better stability, avoiding numerical instability issues in the case of interpolated DI. Furthermore, it exhibits improved computational speedup when applied to region-based DI for the 1D Burgers' equation. WLaSDI significantly enhances the performance of LaSDI in terms of stability, robustness, and efficiency. 

There are several important directions for future research within the context of the WLaSDI framework. First, while our utilization of the weak form has demonstrated improved results, the choice of test function and its support play a crucial role in the accuracy and efficacy of the framework. Consequently, further investigation into the selection of these hyperparameters is necessary. Secondly, our current framework primarily considers the case where parameters only influence the initial conditions. Therefore, the development of parameterizations for factors beyond the initial condition is a promising direction for future work. The current implementation of WLaSDI only discovers a reduced-order model in space. However, in many scenarios, it could be computationally advantageous to explore space-time reduction techniques as part of the compression step. Currently, users are faced with several decisions, including choosing between nonlinear and linear compression, selecting from various local and global techniques, and determining multiple hyperparameters, such as the latent space dimension and the number of training points. It would be advantageous to provide guidelines to aid users in making these decisions along with automating the selection of hyperparameters. As demonstrated in \cite{FriesHeChoi2022ComputerMethodsinAppliedMechanicsandEngineering}, users may currently opt for linear compression when the singular values decay rapidly and rely on the truncation of the cumulative sum of singular values to determine the latent space dimension. Further investigation into this aspect, especially in the case of nonlinear compression, is warranted.

In light of these future directions, WLaSDI provides a fast and robust data-driven reduced-order simulation capability and we play to employ WLaSDI in diverse, high-impact fields such as climate science and fusion energy.


\section*{Declaration of competing interest}
The authors declare that they have no known competing financial interests or personal relationships that could have appeared to influence the work reported in this paper.

\section*{Acknowledgements}
This work was supported in part by Rudy Horne Fellowship to AT. This work also received partial support from the U.S. Department of Energy, Office of Science, Office of Advanced Scientific Computing Research, as part of the CHaRMNET Mathematical Multifaceted Integrated Capability Center (MMICC) program, under Award Number DE-SC0023164 to YC at Lawrence Livermore National Laboratory, and under Award Number DE-SC0023346 to DMB at the University of Colorado Boulder. Lawrence Livermore National Laboratory is operated by Lawrence Livermore National Security, LLC, for the U.S. Department of Energy, National Nuclear Security Administration under Contract DE-AC52-07NA27344. 
This work utilized the Blanca condo computing resource at the University of Colorado Boulder. Blanca is jointly funded by computing users and the University of Colorado Boulder.



\newpage
\clearpage
\bibliographystyle{plain}

\bibliography{./tran01.bib}

\end{document}